\documentclass[pdflatex,sn-mathphys-num]{sn-jnl}

\usepackage{graphicx}%
\usepackage{multirow}%
\usepackage{amsmath,amssymb,amsfonts}%
\usepackage{amsthm}%
\usepackage{mathrsfs}%
\usepackage[title]{appendix}%
\usepackage{xcolor}%
\usepackage{textcomp}%
\usepackage{manyfoot}%
\usepackage{booktabs}%
\usepackage{algorithm}%
\usepackage{algorithmicx}%
\usepackage{algpseudocode}%
\usepackage{listings}%

\usepackage{tabularx}

\usepackage{multicol}
\usepackage{multirow}
\usepackage{tikz}
\usetikzlibrary{arrows,arrows.meta,shapes,positioning,shadows,trees,calc,intersections,external,intersections}
\usetikzlibrary{positioning}
\usetikzlibrary{shapes.geometric, arrows}

\newcommand{\be}{\begin{equation}}
\newcommand{\ee}{\end{equation}}
\newcommand{\bea}{\begin{eqnarray}}
\newcommand{\eea}{\end{eqnarray}}

\newcommand{\inputnum}{3} 
\newcommand{\hiddennum}{4}  
\newcommand{\outputnum}{2} 
\tikzstyle{startstop} = [rectangle, rounded corners, 
    text width=7em, minimum height=1cm,text centered, draw=black, fill=red!30]
\tikzstyle{io} = [trapezium, trapezium left angle=70, trapezium right angle=110, 
    text width=7em, minimum height=1cm, text centered, draw=black, fill=blue!30]
\tikzstyle{process} = [rectangle, 
    text width=7em, minimum height=1cm, text centered, draw=black, fill=orange!30]
\tikzstyle{decision} = [diamond, minimum height=1cm, text centered, text width=5.5em, node distance=3cm, draw=black, fill=green!30]
\tikzstyle{arrow} = [thick,->,>=stealth]

\theoremstyle{thmstyleone}%
\theoremstyle{thmstyletwo}%

\theoremstyle{thmstylethree}%

\begin{document}

\title[Cosmo-Learn: code for learning cosmology using different methods and mock data]{Cosmo-Learn: code for learning cosmology using different methods and mock data}


\author*[1,2]{\fnm{Reginald Christian} \sur{Bernardo}}\email{reginald.christian.bernardo@aei.mpg.de}

\author[3]{\fnm{Daniela} \sur{Grand\'on}}\email{daniela.grandon@uv.cl}

\author[4,5]{\fnm{Jackson Levi} \sur{Said}}\email{jackson.said@um.edu.mt}

\author[3]{\fnm{V\'ictor H.} \sur{C\'ardenas}}\email{victor.cardenas@uv.cl}

\author[6]{\fnm{Gene Carlo} \sur{Belinario}}\email{gcbelinario@nip.upd.edu.ph}

\author[6,7,8]{\fnm{Reinabelle} \sur{Reyes}}\email{reinareyes.rctp@gmail.com}

\affil*[1]{\orgname{Max Planck Institute for Gravitational Physics (Albert Einstein Institute)}, 
\orgaddress{\city{Hannover}, \postcode{30167}, \country{Germany}}}

\affil[2]{\orgname{Asia Pacific Center for Theoretical Physics}, 
\orgaddress{\city{Pohang}, \postcode{37673}, \country{Republic of Korea}}}

\affil[3]{\orgdiv{Instituto de F\'isica y Astronom\'ia}, \orgname{Universidad de Valpara\'iso}, 
\orgaddress{\street{Avenida Gran Bretaña 1111}, \city{Valparaíso}, \country{Chile}}}

\affil[4]{\orgdiv{Institute of Space Sciences and Astronomy}, \orgname{University of Malta}, 
\orgaddress{\postcode{MSD 2080}, \country{Malta}}}

\affil[5]{\orgdiv{Department of Physics}, \orgname{University of Malta}, 
\orgaddress{\postcode{MSD 2080}, \country{Malta}}}

\affil[6]{\orgdiv{National Institute of Physics}, \orgname{University of the Philippines Diliman}, 
\orgaddress{\city{Quezon City}, \postcode{1101}, \country{Philippines}}}

\affil[7]{\orgname{Philippine Space Agency}, 
\orgaddress{\city{Bagumbayan, Quezon City}, \postcode{1800}, \country{Philippines}}}

\affil[8]{\orgdiv{Research Center for Theoretical Physics}, \orgname{Central Visayan Institute Foundation}, 
\orgaddress{\city{Jagna, Bohol}, \postcode{6308}, \country{Philippines}}}








\abstract{
We present \texttt{cosmo\_learn}, an open-source \texttt{python}-based software package designed to simulate cosmological data and perform data-driven inference using a range of modern statistical and machine learning techniques. Motivated by the growing complexity of cosmological models and the emergence of observational tensions, \texttt{cosmo\_learn} provides a standardized and flexible framework for benchmarking cosmological inference methods. The package supports realistic noise modeling for key observables in the late Universe, including cosmic chronometers, supernovae Ia, baryon acoustic oscillations, redshift space distortions, and gravitational wave bright sirens. We demonstrate the internal consistency of the simulated data with the input cosmology via residuals and 
parameter recovery using a fiducial $w$CDM model. Built-in learning and inference modules include traditional Markov Chain Monte Carlo, as well as more recent approaches such as genetic algorithms, Gaussian processes, Bayesian ridge regression, and artificial neural networks. These methods are implemented in a modular and extensible architecture designed to facilitate comparisons across inference strategies in a common pipeline. By providing a flexible and transparent simulation and learning environment, \texttt{cosmo\_learn} supports both educational and research efforts at the intersection of cosmology, statistics, and machine learning.
}




\maketitle

\section{Introduction} \label{sec:intro}

The search for new physics has taken on central importance in meeting the growing challenges of modern cosmology. Before the precision cosmological era, supernovae were found fainter than expected in the cosmological model without dark energy, and the fit is improved by introducing $\Lambda$, which supported the observed accelerated expansion. Now, precision measurements of the expansion of the Universe \cite{SupernovaSearchTeam:1998fmf,SupernovaCosmologyProject:1998vns} show a faster rate than predicted by the $\Lambda$CDM concordance model \cite{DiValentino:2021izs}, when early Universe data sets are considered. As a consequence, there has been an increased effort to produce realistic potential resolutions to this open question. Moreover, the standard concordance model is undergoing a reassessment of the constituent components in its makeup, including the direct observation of cold dark matter (CDM) \cite{Baudis:2016qwx,Bertone:2004pz,Gaitskell:2004gd}, the theoretically problematic cosmological constant \cite{Peebles:2002gy,Copeland:2006wr}, and observational measurements of cosmic inflation \cite{Baumann:2014cja,LiteBIRD:2022cnt}. The latest addition to the family of open questions is that of the Hubble tension \cite{CosmoVerse:2025txj,DiValentino:2020zio}, which expresses the discordance between estimates of the Hubble constant in the local Universe and concordance model based measurements using early-time observations. This effect is compounded by the possibility of tensions in the rate of growth of large scale structures \cite{CosmoVerse:2025txj,DiValentino:2020vvd,Benisty:2020kdt,Poulin:2022sgp,LeviSaid:2021yat,DiValentino:2025sru}. There are several competing interpretations of the question of cosmic tensions \cite{CosmoVerse:2025txj,Abdalla:2022yfr}, but it is progressively becoming a characteristic between direct measurements in the late Universe \cite{Riess:2021jrx,Wong:2019kwg,Anderson:2023aga,Freedman:2020dne}, and inferred cosmological parameter constraints using measurements from the early Universe \cite{Aghanim:2018eyx,DES:2021wwk,eBOSS:2020yzd,Zhang:2021yna,Cooke:2017cwo}. 

As the wealth of data releases continues to grow, the likelihood of the
Hubble tension question being the result of a single systematic that permeates through the data pipelines in question becomes increasingly less likely.
In this context, there has been a spectrum of potential directions beyond $\Lambda$CDM in which to build potential cosmological solutions. These have taken the form of a number of promising expressions including modifications in the physics beyond recombination with early dark energy models \cite{Poulin:2023lkg} being the most promising in this regime, additional extra relativistic degrees of freedom \cite{DiValentino:2021imh}, as well as the plethora of suggestions for modified gravitational physics \cite{Addazi:2021xuf,CANTATA:2021ktz,Bahamonde:2021gfp,Bamba:2012cp, Nojiri:2010wj,Nojiri:2017ncd} in different sectors of phenomenology. This spectrum of different physics at all scales of phenomenology has produced a deluge of potential directions in which to build potentially viable physical models, which has only increased the need for a more standardized benchmark on which to compare cosmologies. 

The vast array of cosmological frameworks on which models are built are in direct competition with each other, and in many cases produce models that are degenerate in terms of their constraints through observational measurements. This leads to an environment where having precisely comparable tests has become an issue of central importance. 

An additional direction is cosmological reconstruction, where the focus shifts from constraining cosmological models to directly inferring the underlying expansion history and growth rate of structure from data. Such approaches aim to remain agnostic to the detailed form of the cosmological model, allowing for a more data-driven view of the Universe. Reconstructions of the Hubble parameter, the dark energy equation of state, or the growth rate of structure provide a flexible way of assessing whether observational data favor the $\Lambda$CDM concordance model \cite{Shafieloo:2009hi,Seikel2012,Yahya:2013xma, Zhao:2017cud}. Cosmological reconstructions serve as a bridge between theoretical model-building and empirical investigations, and they highlight where tensions may arise without relying on model priors. In this sense, reconstructions complement both traditional parameter estimation and the newer statistical and machine learning tools \cite{Bernardo:2021cxi}.

It is in this context that we aim to homogenize some aspects of potential benchmark testing for model comparisons. We do this through a new open and transparent pipeline for mock data generation, where mock data is generated as a representation of different real observational measurements for different classes of
cosmological models.
In this way, mock data around an underlying fiducial cosmology can be built by mimicking the real data noise distribution and assuming a mean from a noisy realization of the underlying cosmology.
This will be important for producing standardized processes for various cosmological scenarios. As a preliminary approach, we include data sets in the local Universe in this first iteration of release. This includes a cosmic chronometer data set release \cite{2010JCAP...02..008S, 2012JCAP...08..006M, 2014RAA....14.1221Z, Moresco:2015cya, Moresco:2016mzx, Ratsimbazafy:2017vga}, the Pantheon+ compilation \cite{Riess:2021jrx,Scolnic:2021amr,Brout:2022vxf}, redshift space distortions measurements \cite{Kazantzidis:2018rnb} related to baryonic acoustic oscillations \cite{DESI:2024mwx}, and gravitational wave bright sirens. This gives a combination of relatively standard and new data sets under which to consider the resulting mock data generation pipeline. Moreover, this gives a healthy product that probes deep into relatively high redshift space. Through this implementation, we hope to provide a community toolkit that can more easily tie into processing pipelines requiring a mock data generation implementation.

{The establishment of an accessible and transparent mock data generator is important to build toolkits for the next generation of cosmological probes given their nuances and individual complexities. As these surveys build momentum toward their eventual data releases, collaborations such as Euclid \cite{Laureijs:2011gra} and the Large Synoptic Survey Telescope \cite{2009arXiv0912.0201L}, among others, stand to benefit from standardized mock data pipelines that allow methods to be developed, tested, and compared in a controlled setting. Moreover, having an easy-to-use mock data generator specific to cosmology will also accelerate efforts to develop toolkits related to other cosmological probes which may involve more nuanced information such as in the cases of weak lensing, galaxy cluster counts or halo occupation distributions, among others.}

Another aspect of the issue of constraining cosmological models in a statistically robust manner is related to the method by which the constraint analysis takes place.
Traditionally, Bayesian inference is performed using Markov chain Monte Carlo (MCMC) methods \cite{Hogg:2017akh,doi:10.1146,2017ARA&A..55..213S}, in which a Markov chain is constructed to approximate the posterior distribution of the model parameters. By generating a large number of correlated samples from this chain, one can approximate credible intervals, and other summary statistics of the posterior.
While effective, this method is computationally expensive both when large data sets are considered, and similarly when complex physical models with multiple nuisance parameters and numerically solved parameters 
are incorporated. In recent years, there has been a concerted effort to meet these challenges through the development of novel statistical tools based on machine learning in order to
understand cosmological models,
as well as to constrain them. {One popular approach has been to use simulation-based inference (SBI) techniques where the outputs of a full numerical simulator are emulated using different methods. These techniques offer an important way to accelerate the running of complex cosmological simulations while retaining numerically precise model predictions. SBIs primarily depend on different neural network configurations which aim to mimic the (1) Bayes posterior direction \cite{Miller:2021hys,2019arXiv190507488G}, (2) the individual likelihood output for each iteration \cite{2018arXiv180507226P,2018arXiv180509294L}, or (3) the marginalized outputs for model parameters \cite{Miller:2022haf,2019arXiv190304057H}. Other approaches exist to accelerate the constraint of complex cosmological models. However, SBIs form a key part of this spectrum of new methods to assess cosmological models. }Another important technique is that of genetic algorithms (GAs) \cite{Medel-Esquivel:2023nov,Bernardo:2025flj,Bernardo:2025zbv,DiValentino:2025sru} where a potentially unbounded space of model realizations can be searched using techniques based on natural selection-inspired approaches. Here, the parameter values define the infinite space of potential realizations, and a potentially faster approach to the best-fit constraint is approximated using the GA method. This may be more effective for the increasingly large space of highly complex cosmological models.

Another key technique in the toolkit of novel methods is that of Gaussian processes (GPs) \cite{10.5555/971143,10.5555/1162254} where an underlying data set is used to train a kernel function, which is a representation of the relationships
between the constituents of the data set. The goal of this approach is to produce a kernel that best represents the data set combination. On the other hand, this produces a powerful way to generate data sets that change the data set distribution while also being genuine representations of the measurements. In the same spirit, Bayesian ridge regression (BRR) \cite{bishop2006pattern,neal2012bayesian} offers a parametric approach to estimating data points and their associated uncertainties for a redshift range. Here, the parametric approach is dynamic in that it can be self-modulated in relation to the complexity of the underlying data sets under consideration. These two methods are closely related to one another.

These methods are powerful in their predictive power but they take on a number of statistical assumptions in their estimation method. In order to counter this potential issue, neural network architectures \cite{Mukherjee:2022yyq,Dialektopoulos:2021wde,Benisty:2022psx,Escamilla-Rivera:2021rbe, Grandon:2022gdr} offer a stronger statistically robust base on which to produce representative data sets that mimic the underlying real measurements. Neural network methods are based on a biologically-inspired approach in which neurons are organized into layers with connections between neurons across different layers. These connections and the neuron responses are controlled by non-physical hyperparameters. Through this approach, a realistic estimation of the data point values and associated uncertainties can be directly obtained in a natural way.

In this work, we explore the question of how these methods compare against each other. We do this using different combinations of mock generated data in order to give a clear benchmark on which to probe these techniques. In this way, we aim to extend the discussion in the literature on possible novel methods on which to rely beyond traditional approaches. In Section~\ref{sec:simulating_cosmological_data}, the mock data generation approach is outlined, as are the observed data sets and the cosmological model on which this is based. The validation approach is then outlined in Section~\ref{sec:testing_cosmo_learn} where $w$CDM is used to estimate cosmological parameters, which acts as the underlying fiducial model. Then, in Section~\ref{sec:learning_methods} the methods outlined here are applied to the mock data sets in order to reconstruct the cosmic evolution deep into intermediate redshift space. This reconstruction of the expansion and large scale structure formation may provide key insights in disentangling the degeneracies between cosmological models and understanding nuances in the comparison of their underlying physics.

The \texttt{cosmo\_learn} package 
can be found at \href{https://github.com/reggiebernardo/cosmo_learn}{https://github.com/reggiebernardo/cosmo\_learn}; which come with tutorial notebooks for an interactive and easier familiarization of the modules. This uses the modules \texttt{astropy} \cite{Astropy:2013muo, Astropy:2018wqo, Astropy:2022ucr}, \texttt{scikit-learn} \cite{scikit-learn}, \texttt{reFANN} \cite{Wang:2019vxv} (ANN based on \texttt{pytorch} \cite{Paszke:2019xhz}), \texttt{gwcatalog} \cite{Ferreira:2022jcd, Ferreira:2023tat, Ferreira:2023awf}, \texttt{emcee} \cite{emcee}, \texttt{geneticalgorithm}, as well as \texttt{numpy} \cite{harris2020array}, \texttt{scipy} \cite{2020SciPy-NMeth}, \texttt{matplotlib} \cite{Hunter:2007}, and \texttt{corner} \cite{corner}.

\section{Simulating cosmological data}
\label{sec:simulating_cosmological_data}

In this section, we present the default setup used by \texttt{cosmo\_learn} to generate a mock data realization based on a provided set of authentic samples and a `true' cosmology. Then, we describe in brief each of the mock data sets that can be simulated with \texttt{cosmo\_learn}: SNe Type Ia distance moduli, CC, BAO, RSD growth rate measurements, and GW standard sirens' luminosity distances.

\subsection{Mock data generation}
\label{subsec:mock_data_generation}

In \texttt{cosmo\_learn}, mock data $\left(z_{\rm mock}, y_{\rm mock} \pm \Delta y_{\rm mock} \right)$ is \textit{by default} generated based on actual observed samples $\left(z_{\rm obs}, y_{\rm obs} \pm \Delta y_{\rm obs}\right)$ and a `true' cosmology $y_{\rm true}[H_0, {\mathbf \Omega}]$ in the following steps:
\begin{enumerate}
    \item Fix the redshifts to match actual ones, $z_{\rm mock} = z_{\rm obs}$;
    \item Draw the mean from normal distribution around the true cosmology with a variance fixed by observation, $y_1 = {\cal N} \left( y_{\rm true}, \left( \Delta y_{\rm obs} \right)^2 \right)$;
    \item Final mock data $\left(z_{\rm mock}, y_{\rm mock} \pm \Delta y_{\rm mock} \right)$ is $\left(z_{\rm obs}, y_1 \pm \Delta y_{\rm {obs}} \right)$.
\end{enumerate}
This default setting leads to a conservative mock data realization to avoid possible spurious forecasts. For the purposes of demonstration and entry for students, we found that this is sufficient. Experts on the other hand can modify the function \texttt{how\_to\_mock} in \texttt{cosmo\_learn.py} to play with the mock data generation, such as consider a full covariance $\Sigma$ and draw the redshift samples $z_{\rm mock}$ using an empirical Gamma {or Beta} distributions \cite{Dialektopoulos:2021wde, Ferreira:2022jcd}.

The present version `true' cosmology is based on $w$CDM (Section \ref{subsec:wcdm}). Thus, an instance of \texttt{cosmo\_learn} can be initiated by providing a value of the Hubble constant $H_0$, nonrelativistic matter density parameter $\Omega_{m0}$, the smoothed amplitude of matter fluctuations $\sigma_8$, and the dark energy equation of state $w$. It is useful to provide a seed specific to an instance or realization of mock data (in a simulated universe) for reproducibility. A minimal example to start an instance of \texttt{cosmo\_learn} with mock data generated by the default setting is
{\small \begin{verbatim}
my_cl=CosmoLearn([H0, Om0, w0, s8], seed=14000605)
my_cl.make_mock(mock_keys=[`CosmicChronometers', `SuperNovae', `BrightSirens'])
\end{verbatim}}
\noindent where the $H_0$ is in units of km s Mpc$^{-1}$ and $\Omega_{m0}, w_0, S_8$ are dimensionless; e.g., \cite{DESI:2024mwx, Aghanim:2018eyx}
\begin{verbatim}
H0, Om0, w0, s8 = 67.74, 0.3095, -0.997, 0.834
\end{verbatim}
The above two lines generate an instance of \texttt{cosmo\_learn} in universe 14000605 with a true cosmology given by the DESI Year 1 Early Data Release results \cite{DESI:2024mwx}, with cosmic chronometers, supernovae, and GW bright sirens mock data. Obviously, the choice of true cosmology based on a set of cosmological parameters $(H_0, \Omega_{m0}, w, S_8)$ is arbitrary for learning purposes and completely up to the user; where $S_8=\sigma_8 \sqrt{\Omega_{m0}/0.3}$. Keys \begin{verbatim} `BaryonAcousticOscillations', `RedshiftSpaceDistorsions' \end{verbatim} for the method \texttt{make\_mock} are similarly available to generate baryon acoustic observation samples and redshift space distortion growth rate measurements.

\begin{figure}[h!]
    \centering
    \includegraphics[width=0.9\textwidth]{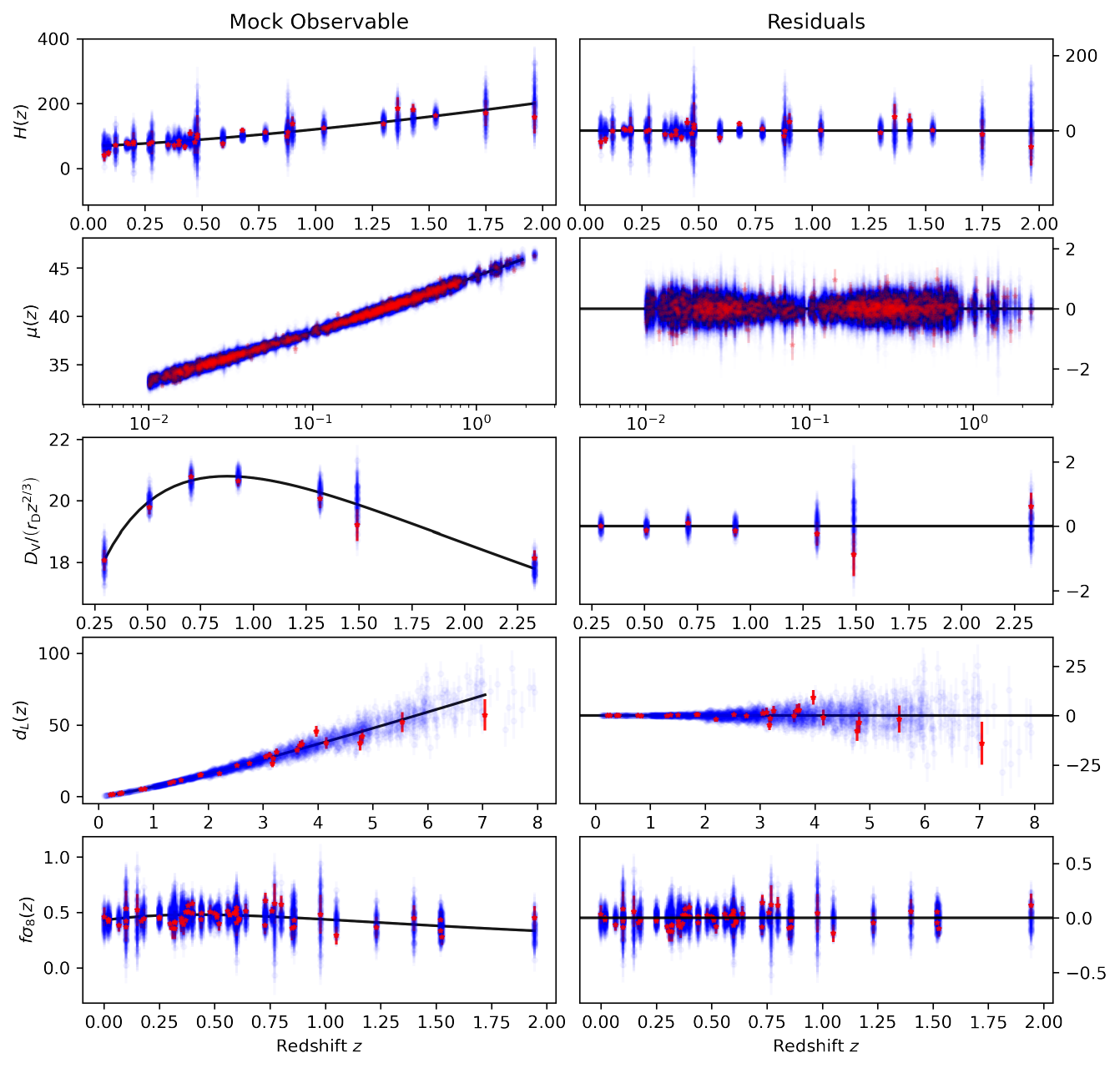}
    \caption{Superimposed simulated data and their residuals across realizations in \texttt{cosmo\_learn} (blue, seeds 1-100) and the sampled mock data in universe 14000605 (red). The `true' cosmology is represented by the black solid lines.}
    \label{fig:simulated_data}
\end{figure}

Mock data and their corresponding residuals across 100 realizations are shown in Figure \ref{fig:simulated_data}. This shows that the generated data across multiple realizations are centered on the true cosmology, i.e., residuals are white, as expected since they were drawn from a Gaussian distribution. Mock data on a single realization, e.g., reference universe 14000605, can manifest $\sim$1-2$\sigma$ deviations due to the noise in the actual data. Later tests (Section \ref{sec:testing_cosmo_learn}) will show that cosmological parameters can be inferred with $\sim$1-2$\sigma$ deviations relative to the true cosmology, consistent with expectations.

Mock data generated in \texttt{cosmo\_learn} are divided into a training set (90\% of the full data) and a test set (10\%) that are specific to the random state assigned per realization. This is used for training the methods integrated into the package (Section \ref{sec:learning_methods}).

\subsection{Standard cosmological data}
\label{subsec:standard_cosmological_data}

We describe the data sets considered in \texttt{cosmo\_learn} for generating mock data. The package is built to be versatile and allow additional and other data sets to be added as needed. Below, we discuss the built-in data sets that the package uses, which are centered in the local Universe, but which can be extended.

\subsubsection*{Cosmic Chronometers}

Cosmic Chronometers (CC) provide a practical way to directly constrain the Hubble expansion rate $H(z, \Theta)$ at different redshifts due to the differential aging method on which these data sets are built \cite{Moresco:2024wmr}. In our implementation, we adopt 31 data points as compiled from Ref.~\cite{2014RAA....14.1221Z,Jimenez:2003iv,Moresco:2016mzx,Simon:2004tf,2012JCAP...08..006M,2010JCAP...02..008S,Moresco:2015cya}. The spectroscopic dating technique used in this method expresses the differential aging of cosmological local galaxies with their redshifts, $\Delta z / \Delta t$, so that the Hubble expansion at those redshifts can be determined $H_{\mathrm{obs}}(z) = -(1+z)^{-1} \Delta z/ \Delta t$. In this way, CC data is a reliable source of data points that are free of cosmological models, and moreover their direct nature renders them independent of any complex calculations. Their only dependence is on the absolute aging of galaxies which is a fairly well understood feature in astrophysics \cite{Jimenez:2001gg}.

\subsubsection*{Supernovae Type Ia}

The Pantheon+ ($\mathrm{PN}^+$\,\&\,SH0ES) sample \cite{Riess:2021jrx,Scolnic:2021amr,Brout:2022vxf} contains one of the largest samples of type Ia supernovae (SNe) with 1701 measurements spanning a redshift range of $0.0$ to $2.26$. The Pantheon+ analysis uses Cepheid hosts as distance anchors together with geometric calibrators. In this approach, the degeneracy between the absolute magnitude $M$ and $H_0$ is effectively addressed through the use of SH0ES prior information. The expansion rate at different redshifts, $H(z, \Theta)$, is measured through the observed apparent magnitude, $m$, through the relation
\begin{equation} \label{eq:dist_mod}
    \mu(z_i, \Theta) = m - M = 5 \log_{10}[d_L(z_i, \Theta)] + 25 \,,
\end{equation}
where $d_L(z_i, \Theta)$ represents the luminosity distance to each redshift expressed as
\begin{equation} \label{eq:lum_dist}
    d_L(z_i, \Theta) = c(1+z_i) \int_0^{z_i} \frac{dz'}{H(z', \Theta)} \,, 
\end{equation}
where $c$ is the speed of light. We consider the subset $z > 0.01$ to be conservative of potential issues with very low redshift samples \cite{Pasten:2023rpc, Perivolaropoulos:2023iqj}.

\subsubsection*{RSD Growth Data}

Redshift-space distortions (RSD) offer a powerful observational handle on the large-scale structure of the Universe by measuring anisotropies in the galaxy distribution; caused by peculiar velocities superimposed on the Hubble flow. These anisotropies carry crucial information about the growth of cosmic structures and allow us to track the evolution of the linear matter density contrast, $\delta(z)$, across redshift. Importantly, $\delta(z)$ is not only a tracer of structure formation but also a sensitive probe of gravity on cosmological scales. At the linear perturbation level, the evolution of the density contrast is governed by the differential equation
\begin{equation}
\label{eq:delta_evolution}
    \delta''(a) + \left( \dfrac{3}{a} + \dfrac{H'(a)}{H(a)} \right) \delta'(a) - \dfrac{3}{2} \dfrac{ \Omega_{m0} \, G_{\rm eff}(a, k)/G_{\rm N} }{ a^5 H(a)^2 / H_0^2 } \delta(a) = 0 \,,
\end{equation}
where $G_{\rm eff}(a, k)$ characterizes potential modifications to the gravitational strength on large scales \cite{Silveira:1994yq, Percival:2005vm, Lee:2009gb, 2011JCAP...10..010B, Nesseris:2015fqa, Kazantzidis:2018rnb}. In GR, $G_{\rm eff}(a, k)/G_{\rm N} = 1$, while departures from unity can signal deviations from Einsteinian gravity. Hence, by constraining the evolution of $\delta(z)$ through RSD data, one effectively places bounds on gravity itself at linear cosmological scales.

Observationally, this information is captured through measurements of $f\sigma_8(z)$, the product of the linear growth rate $f(z) = d\ln \delta(z)/d\ln a(z)$ and the root-mean-square matter fluctuation amplitude $\sigma_8$ on scales of $8\,h^{-1}\,\text{Mpc}$. This quantity effectively encapsulates both the rate of structure growth and the normalization of the matter power spectrum.

In the standard cosmological model, the linear matter overdensity evolves as \cite{Nesseris:2015fqa}:
\begin{equation}
\label{eq:matterpert_lcdm}
    \delta(a) = a \,_2F_1\left( \dfrac{1}{3},1, \dfrac{11}{6}; a^3 \left(1 - \Omega_{m0}^{-1}\right)  \right),
\end{equation}
where $\,_2F_1(a,b,c;x)$ is the hypergeometric function. This expression admits generalization to models with a constant dark energy equation of state, a.k.a. $w$CDM \cite{Nesseris:2015fqa}.

The growth rate observable is then given by \cite{LeviSaid:2021yat}:
\begin{equation}
\label{eq:growth_lcdm}
    f\sigma_8(z) = -\sigma_8 (1 + z) \dfrac{\delta'(z)}{\delta(z = 0)}.
\end{equation}
Here, $\sigma_8(z) = \sigma_8 \delta(z)/\delta(z=0)$ reflects the redshift evolution of the fluctuation amplitude.

For our analysis, we employ the compilation of RSD measurements from \cite{Kazantzidis:2018rnb}.

\subsubsection*{Baryon Acoustic Oscillations}

Baryonic acoustic oscillations (BAO) map out the fluctuations of the density profile of visible matter. These repeated oscillations can be used as a standard ruler to precisely measure cosmological distances. In this work, the Dark Energy Spectroscopic Instrument (DESI) data release 1 (DR1) \cite{DESI:2024mwx} was used, which consists of BAO distance scales binned in several redshifts using different tracers, namely the bright galaxy sample (BGS) at $0.1<z<0.4$, the luminous red galaxy (LRG) sample at $0.4<z<0.6$ and $0.6<z<0.8$, a combination of LRG and emission line galaxies (ELG) at $0.8<z<1.1$, the ELG sample at $1.1<z<1.6$, the quasar sample at $0.8<z<2.1$, and the Lyman-alpha forest sample at $1.77<z<4.16$. These seven redshift bins make up the DESI DR1 sample. 

For the BAO DESI DR1 sample, we compute the Hubble distance $D_H(z)$ through the expression $D_H(z) =c/H(z)$, which gives the effective distance along the line-of-sight. The effective distance transverse to the line-of-sight is given by
\begin{equation}
    D_M(z) = c \int^z_0 \frac{dz'}{H(z')} \,,
\end{equation}
which is also related to the angular diameter distance by $D_A = (1+z)^{-1} D_M(z)$. The volume average distance is then given by
\begin{equation}
    D_V(z) = (1+z)^{-1} \left[D_M(z)^2 \frac{cz}{H(z)}\right]^{\frac{1}{3}} \,. 
\end{equation}

BAO measurements are inherently degenerate with the comoving sound horizon at the drag epoch, $r_{\rm D}$, which acts as the standard ruler. This scale corresponds to the distance a sound wave can travel in the primordial photon-baryon fluid before recombination, and is given by
\begin{equation}
    r_{\rm D} = \int_{z_{\rm D}}^\infty \frac{c_s(z)}{H(z)} \, dz,
\end{equation}
where $z_{\rm D} \simeq 1059.94$ is the redshift of the drag epoch \cite{Planck:2018vyg}, and $c_s(z)$ is the sound speed of the baryon-photon fluid. BAO observables are typically reported as ratios such as $D_H(z)/r_{\rm D}$, $D_M(z)/r_{\rm D}$, or $D_V(z)/r_{\rm D}$, ensuring that model predictions can be directly compared with observational data.

\subsubsection*{GW Bright Sirens}

GWs are tidal ripples in spacetime caused by the dynamics of massive astrophysical objects and can manifest as strains that are captured by large-scale interferometers such as the ground-based LIGO-Virgo-Kagra (LVK) detectors and the anticipated space-borne Laser Interferometer Space Antenna (LISA). GWs may originate from the coalescence of compact binaries, rapid rotation of astrophysical bodies, and even phase transitions from the early universe \cite{Sesana:2004gf, Sesana:2007sh, Sathyaprakash:2009xs, Caprini:2009yp, Caprini:2015zlo, Tamanini:2016, LISA:2017pwj, Caprini:2018mtu, LISACosmologyWorkingGroup:2022jok}. 
The signals from these events carry direct information on the luminosity distance $D_L(z, \Theta)$ to the
source, making them powerful standard sirens \cite{Holz:2005df}.
For coalescing compact binaries, the luminosity distance
can be inferred from the strain measurements captured by interferometric GW detectors. If the merger event is accompanied by an electromagnetic (EM) counterpart (i.e. quasar-like luminous flares, radio jets, flares), the redshift $z$ of the source can be independently determined, either spectroscopically or photometrically \cite{Tamanini:2016}. These joint detections, where both GW and EM signals are observed, are referred to as bright standard sirens, and they offer a direct probe of the cosmological expansion through the distance-redshift relation \eqref{eq:lum_dist}. 

At leading Newtonian order, the GW strain amplitude $h(\tau)$ is related to the luminosity distance by
\begin{align}
    h(\tau) \propto \frac{\mathcal{M}^{5/3}}{D_L(z, \Theta)} f(\tau)^{2/3},
\end{align}
where $\mathcal{M}$ is the chirp mass of the binary system, $f(\tau)$ is the instantaneous GW frequency, $h(\tau)$ is the strain, and $\tau$ is the merger time of the binary \cite{Sathyaprakash:2009xs, Holz:2005df}.
Note that merger time scales are typically much shorter than cosmological time scales, even for massive black hole binaries. Since all these quantities are measurable, the luminosity distance can be inferred directly. When the redshift is also available through EM observations, one can construct the Hubble diagram using bright standard sirens in an analogous manner to SNe Ia, following \eqref{eq:lum_dist}.

\begin{figure}[h!]
    \centering
    \includegraphics[width=0.6\textwidth]{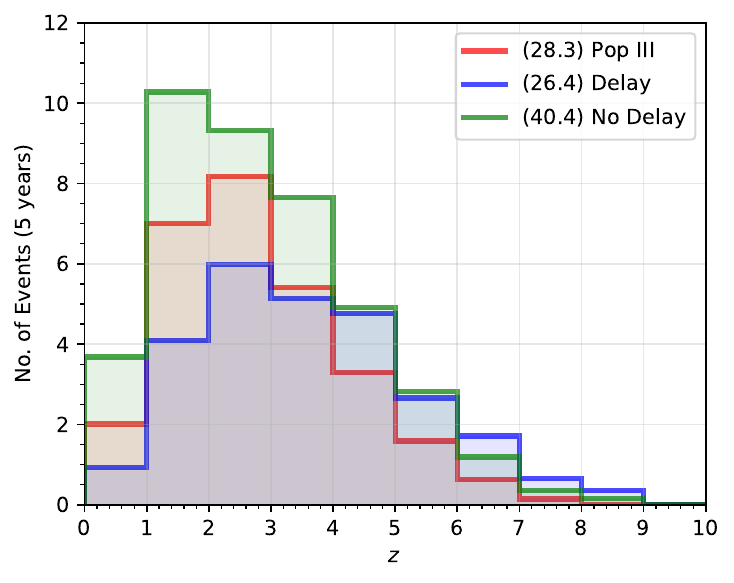}
    \caption{Redshift distribution of GW bright siren events \cite{Ferreira:2022jcd} for different binary population models assuming the eLISA configuration L6A2M5N2 \cite{Tamanini:2016uin}. Each redshift bin corresponds to a particular number of events detected. Numbers in the legend denote the average number of detections over all redshift bins in a 5-year mission lifetime. }
    \label{fig:redshift_dist_sts}
\end{figure}

To construct a mock catalog of standard siren events, we adopt the redshift distribution shown in Figure \ref{fig:redshift_dist_sts}, corresponding to one of the most promising eLISA mission configurations studied in \cite{Tamanini:2016uin}. The figure reflects the expected number of detectable standard sirens with EM counterparts, based on simulations incorporating the eLISA sensitivity curve \cite{Tamanini:2016uin,LISA:2017pwj}. The redshift distribution varies according to the underlying binary population model.

To simulate the events, we first interpret the redshift histogram as a probability distribution function that is nonzero only in the interval $z \in (0.1, 9)$. For each event, we draw a trial redshift from a uniform distribution, and then draw a trial event count from another uniform distribution between the minimum and maximum values of a predetermined expected number of events or mission lifetime; as a default, we consider a 5-year lifetime. The redshift is accepted if the sampled count is less than the maximum expectation; otherwise, the process is repeated. This rejection sampling is continued until the desired number of events for a given binary model is reached. Each accepted redshift is then associated with a source luminosity distance via \eqref{eq:lum_dist}.

To make the mock catalog more realistic, we account for anticipated observational uncertainties to each event \cite{Ferreira:2022jcd, Ferreira:2023tat, Ferreira:2023awf}. We incorporate weak lensing uncertainties using the fit from \cite{Tamanini:2016uin, Hirata:2010ba},
\begin{align}
    \frac{\sigma_{\text{lens}}(z, \Theta)}{d_L(z, \Theta)} \sim 0.066 \left(\frac{1-(1+z)^{-0.25}}{0.25}\right)^{1.8} \left[1-\frac{0.3}{\pi/2}\arctan \left(\frac{z}{z^*}\right)\right],
\end{align}
with $z^* = 0.073$ as used in \cite{Shapiro:2009sr} to control the growth of lensing-induced scatter. We also include uncertainties due to peculiar velocities $v$ following \cite{Tamanini:2016uin, Kocsis:2005vv}:
\begin{align}
    \frac{\sigma_{v}(z, \Theta)}{d_L(z, \Theta)} \sim \left[1 + \frac{c(1+z)^2}{H(z, \Theta)d_L(z, \Theta)}\right] \frac{\sqrt{\langle v^2 \rangle}}{c}.
\end{align}
Then, the instrumental noise from LISA is approximated to scale linearly with distance as described in \cite{tgf_li}:
\begin{align}
    \frac{\sigma_{\text{LISA}}(z, \Theta)}{d_L(z, \Theta)} \sim 0.05 \left(\frac{d_L(z, \Theta)}{36.6 \ \text{Gpc}}\right).
\end{align}
Finally, we account for redshift measurement uncertainties from the EM counterpart. For redshifts below $z\sim2$, which are typically measured spectroscopically, the uncertainties are negligible. For $z \gtrsim 2$, where photometric methods dominate, we use the estimate from \cite{Dahlen:2013fea}:
\begin{align}
    \sigma_{\text{photo}}(z, \Theta) \sim 0.03(1 + z).
\end{align}
These sources of error are added in quadrature to obtain the total uncertainty:
\begin{align}
    \sigma(z, \Theta) = \sqrt{\sigma_{\text{lens}}^2(z, \Theta) + \sigma_{v}^2(z, \Theta) + \sigma_{\text{LISA}}^2(z, \Theta) + \sigma_{\text{photo}}^2(z)}.
\end{align}

We sample the luminosity distance from a Gaussian distribution $\mathcal{N}(d_L(z, \Theta), \sigma(z, \Theta))$ centered at the theoretical $d_L(z, \Theta)$ with standard deviation $\sigma(z, \Theta)$. This procedure yields a full mock standard siren catalog, which can be combined with simulated data from other cosmological probes. Note that the errors explicitly depend on the cosmological parameters. A sample realization is shown in Figure \ref{fig:simulated_data}.

This module in \texttt{cosmo\_learn} is based on our modified version of the \texttt{gwcatalog} codebase \cite{Ferreira:2022jcd, Ferreira:2023tat, Ferreira:2023awf}. We consider a default Pop III binary population model; users are free to change this via the attribute \texttt{pop=`Pop III'} to the \texttt{cosmo\_learn} instance; e.g., key in \texttt{my\_cl.pop\_model=`Delay'} or \texttt{my\_cl.pop\_model=`No Delay'} for the the delay and no delay models, respectively. The mission lifetime can be changed using option \texttt{years=5} when calling \texttt{make\_mock}.

\subsection{\texorpdfstring{$w$}{}CDM cosmology}
\label{subsec:wcdm}

The $w$CDM model is the simplest phenomenological extension of the standard $\Lambda$CDM model, obtained by generalizing the equation of state of dark energy from a fixed value ($w = -1$) to a free parameter $w$, 
\begin{equation}
    p_{\text{DE}} = w \, \rho_{\text{DE}},
\end{equation}
where $w \in \mathbb{R}$ is constant in time.
The value $w = -1$ recovers the $\Lambda$CDM scenario, while $w < -1$ and $w > -1$ correspond to phantom and quintessence-like behaviors \cite{Linder:2002et}, respectively. The Hubble expansion rate in this model is given by
\begin{equation}
    H^2(z) = H_0^2 \left[ \Omega_m (1+z)^3 + \Omega_r (1+z)^4 + \Omega_{\text{DE}} (1+z)^{3(1+w)} \right],
\end{equation}
where $\Omega_{\text{DE}} = 1 - \Omega_m - \Omega_r$ under the assumption of spatial flatness, and the usual scaling laws are assumed for matter ($\propto (1+z)^3$) and radiation ($\propto (1+z)^4$).

The $w$CDM model is frequently used in cosmological analyses as a minimal benchmark to assess deviations from $\Lambda$CDM, particularly in testing the robustness of parameter estimation methods and exploring the sensitivity of current and future surveys to the properties of dark energy. Because it involves only one additional parameter at the background, it provides a controlled framework for quantifying the impact of relaxing the cosmological constant assumption without invoking a full scalar field model or modified gravity. Furthermore, for linearized perturbations, $w$CDM admits an analytical solution for the density contrast of nonrelativistic matter \cite{Lee:2009gb, Nesseris:2015fqa}, allowing for numerically efficient computation of observables depending on linear perturbations such as the growth rate $f \sigma_8(z)$ \cite{Kazantzidis:2018rnb}. Few other cosmological models have similar analytical expressions available, making the $w$CDM model a suitable basis for studying cosmology through numerical analysis.

The $w$CDM model has been extensively employed in data analyses from major cosmological surveys, such as \textit{Planck}~\cite{Planck:2018vyg}, the Dark Energy Survey (DES)~\cite{DES:2021wwk}, and the Pantheon compilation of type Ia supernovae~\cite{Scolnic:2021amr}. These studies have placed tight constraints on $w$, with current data being consistent with $w = -1$ at the few-percent level, though future surveys are expected to improve these bounds significantly.

\texttt{cosmo\_learn} uses the analytical expressions for the background and linear perturbation in $w$CDM in \cite{Nesseris:2015fqa}. By default, DE perturbations are turned off through the attribute \texttt{de\_model=`no pert'}. This can be changed to one with DE perturbations but with the sound speed, $c_s$, fixed to zero as \texttt{de\_model=`static'}. Alternatively, DE perturbations can be turned on with a nonzero sound speed by fixing \texttt{de\_model=`dynamic'} and configuring the attribute \texttt{k2cs2=1e-10}, corresponding to the product $k^2 c_s^2$ for modes with a wavenumber $k$. For further details, see \cite{Nesseris:2015fqa}.

\section{Testing mock data}
\label{sec:testing_cosmo_learn}

\begin{figure}[h!]
\centering
\includegraphics[width=0.9\linewidth]{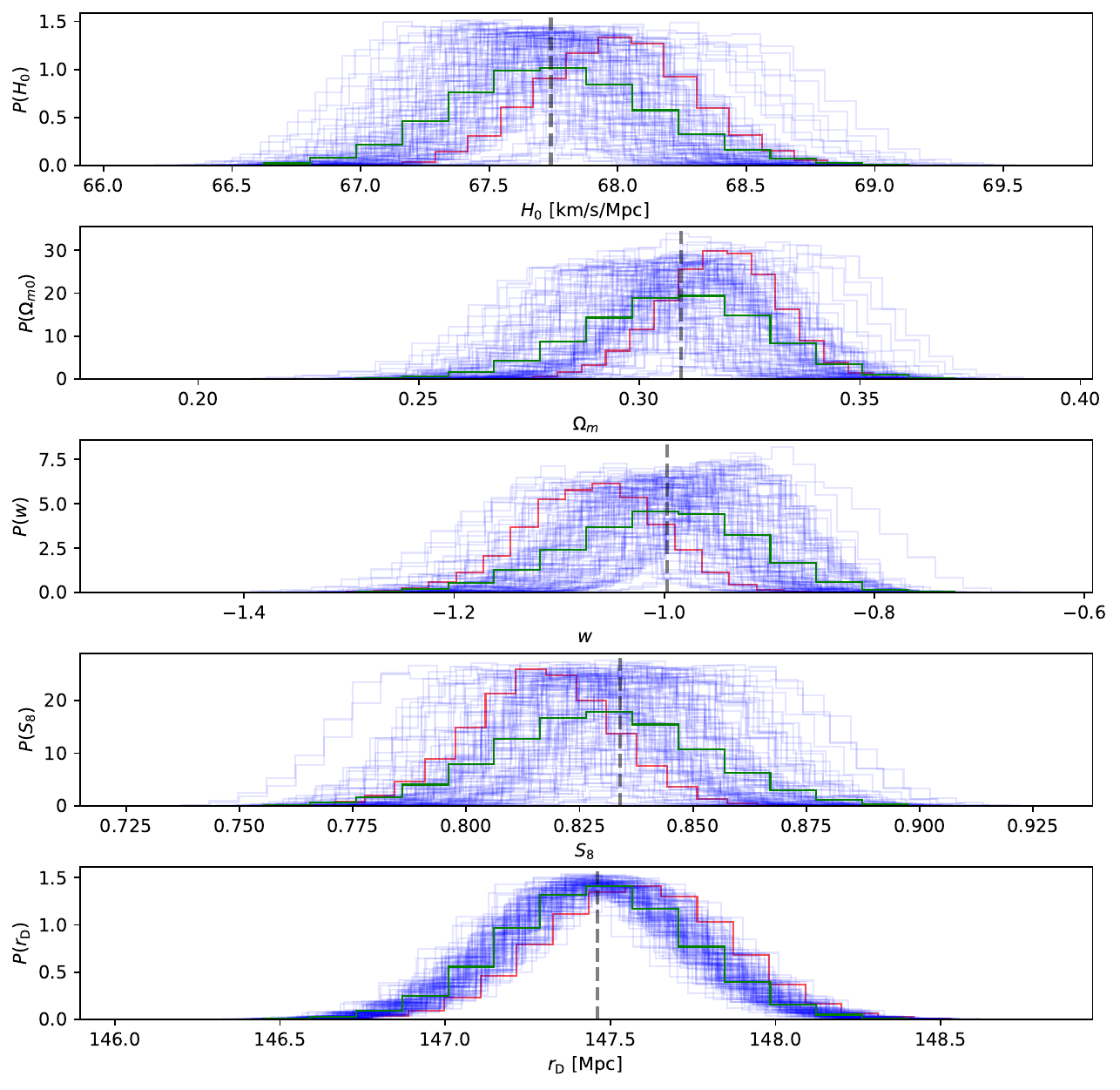}
\caption{Histograms of inferred cosmological parameters from MCMC runs across 100 mock realizations (blue) and for a representative realization, the reference universe 14000605 (red). Gray vertical lines indicate the true cosmology. Green histogram shows the posterior stacked across realizations is a Gaussian centered on the true cosmology. }
\label{fig:constraints}
\end{figure}

To validate the mock data generation beyond the appearance of white residuals, we demonstrate that the input cosmology can be statistically recovered through parameter inference. Specifically, we use the built-in MCMC engine in \texttt{cosmo\_learn} which leverages \texttt{emcee} to estimate the posterior distributions of cosmological parameters under a $w$CDM model. The resulting constraints, aggregated over 100 mock realizations, are shown in Figure~\ref{fig:constraints}, where all available observables (Section \ref{subsec:standard_cosmological_data}) are utilized.

The constraints from individual realizations such as the reference universe demonstrate that the input cosmological parameters are recoverable within $\sim$1–2$\sigma$, depending on the noise level. Furthermore, when posteriors from all realizations are stacked, the resulting ensemble distribution closely approximates a Gaussian centered on the true cosmology. This supports the conclusion that the residuals are effectively white when analyzed over many mock data sets (see Figure~\ref{fig:simulated_data}).

A particularly noteworthy result is the precise constraint on the sound horizon at the drag epoch, $r_{\rm D}$, which shows sub-$1\sigma$ deviations on average. This parameter remains fixed for a given instance of \texttt{cosmo\_learn} and is unaffected by noise, even though it must still be sampled during BAO-based inference.

These MCMC-based results confirm that the mock data generated by \texttt{cosmo\_learn} are statistically consistent with the input cosmology. In the next section, we describe additional inference tools beyond MCMC that are integrated into the package for flexibility and benchmarking.

\section{Learning Methods} \label{sec:learning_methods}
We describe learning methods in \texttt{cosmo\_learn} (MCMC, GA, GP, BRR, ANN) that users can readily play with to seamlessly perform parameter estimation and cosmological reconstruction. 

\subsection{MCMC} \label{subsec:mcmc}

Markov chain Monte Carlo methods (MCMC) \cite{Hogg:2017akh,doi:10.1146,2017ARA&A..55..213S} has become a core tool in estimating posterior parameter distributions for cosmological models in Bayesian inference studies. In this framework, a Monte Carlo approach is taken wherein a large number of random samples are drawn from a normal distribution. The parameters that make up each draw are connected to physical predictions through a system of representative equations such as the Friedmann or Boltzmann equations. The MCMC pipeline also necessitates a set of observational data to be used in order to compare the system predictions in the Monte Carlo segment. As the number of random draws for a sample increases, the parameter means contained in the sample gradually asymptote to the underlying means. This is the benefit of the approach since calculating the mean directly from a system of equations can be cumbersome for most scenarios, while the Monte Carlo method can be used more directly in a vast array of scenarios. The addition of the Markov chain property to the analysis pipeline embeds the idea that the random draws for each sample iteration are connected together in a sequence such that successive draws will either be improved samples in the sequence or discarded in the chain. Notice the important property that this process only requires information on the previous sample iteration and not any of the elements before that.

This approach is particularly useful for Bayesian inference \cite{2013PASP..125..306F,asensio_ramos_arregui_2018,2008ConPh..49...71T} analyses where posterior distributions are often times difficult to determine. Using Bayes' theorem, for a data set $X_{\rm Obs}$ the inferred posterior distribution will be defined by
\begin{equation}
    p(\Theta | X_{\rm Obs}) = \frac{L(\Theta | X_{\rm Obs}) p(\Theta)}{\int L(\Theta' | X_{\rm Obs}) p(\Theta') d\Theta'}\,,
\end{equation}
where $L(\Theta | X_{\rm Obs})$ represents the likelihood of predicting the data given a physical model with parameter $\Theta$, called the likelihood function, $p(\Theta)$ is the prior distribution which normally represents a Crenel function on the parameters that make up the underlying physical model. The denominator is called the marginal likelihood and denotes the likelihood produced by the data itself. This probability of evidence likelihood integral is at the core of the MCMC estimator approach. Naturally, one would like to find the posterior distribution of the parameters given the data. To achieve this, the MCMC pipeline takes successive iterations of $\Theta'$ to estimate the averaged sum approximation of this integral. The end result is that the marginal likelihood is approximated by the mean for $n$ samples drawn giving $(1/n) \Sigma_i L(\Theta' | X_{\rm Obs})$ for each drawn $\Theta'$. Through these successive iterations, the final goal is to determine the posterior distributions of the parameters in $\Theta$ given the observational data $X_{\rm Obs}$. \\

The results of MCMC over a 100 realizations are shown in Figure \ref{fig:constraints}. In \texttt{cosmo\_learn}, MCMC in a realization can be performed as follows (after initializing a \texttt{cosmo\_learn} instance \texttt{my\_cl} and mock data):
{\fontsize{8}{10}\selectfont \begin{verbatim}
# initialize wcdm likelihood
llprob=lambda x: my_cl.llprob_wcdm(x, prior_dict=prior_dict, rd_fid_prior=rd_fid_prior)

# setup and perform mcmc
p0=[70, 0.3, -1, 0.8, 147] # initial point
dres=[0.05, 0.005, 0.01, 0.01, 0.005] # resolution
nwalkers=15; nburn=100; nmcmc=2000 # number of walkers, burn in, steps per walker
my_cl.get_mcmc_samples(nwalkers, dres, llprob, p0=p0, nburn, nmcmc)
\end{verbatim}}

\begin{figure}[h!]
    \centering
    \includegraphics[width=0.9\textwidth]{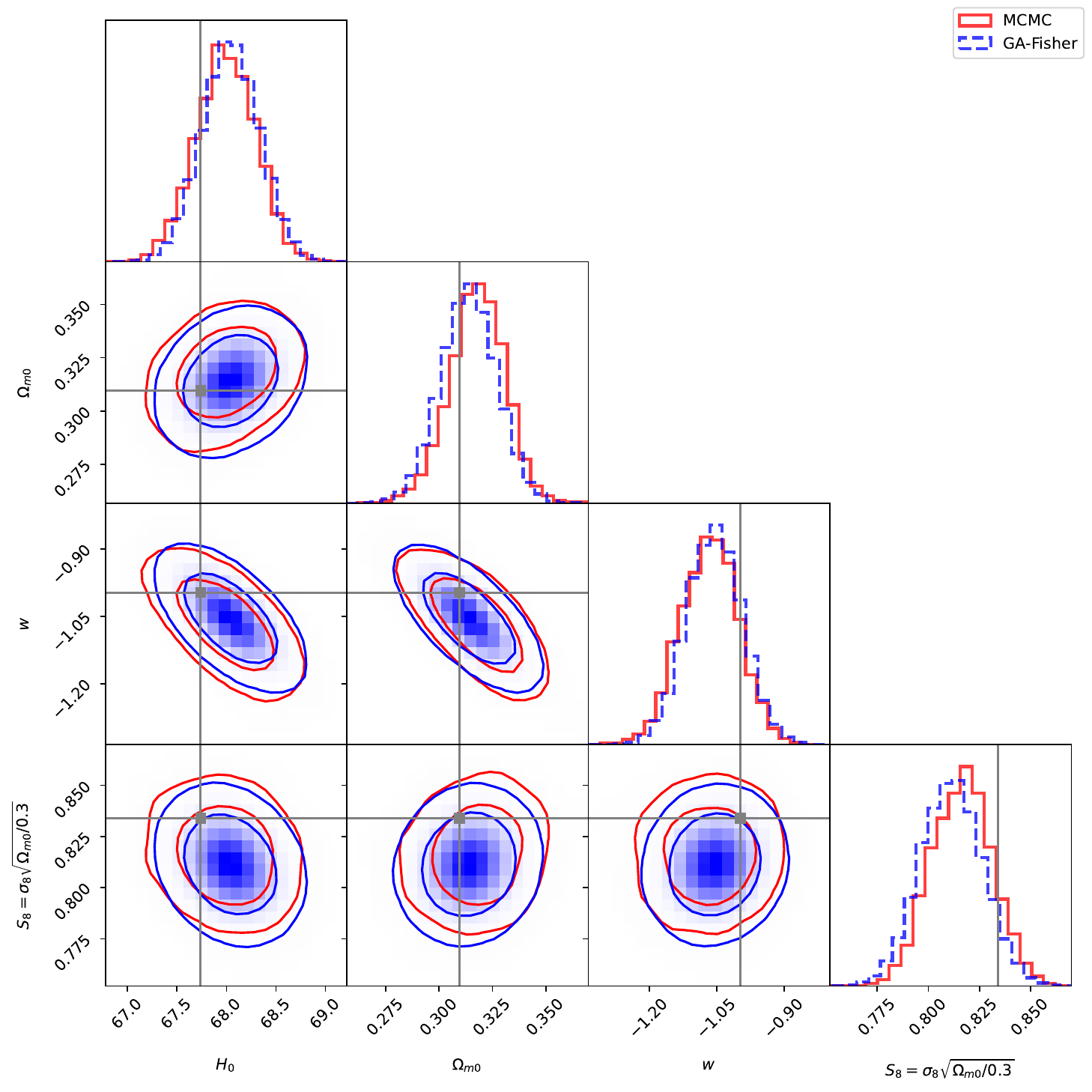}
    \caption{MCMC (red) and GA-Fisher (blue) parameter estimates in $w$CDM cosmology in the reference realization (seed/universe 14000605). Gray horizonal and vertical lines show the true cosmology.}
    \label{fig:corner_ga_fisher}
\end{figure}

\begin{figure}[h!]
    \centering
    \includegraphics[width=0.9\textwidth]{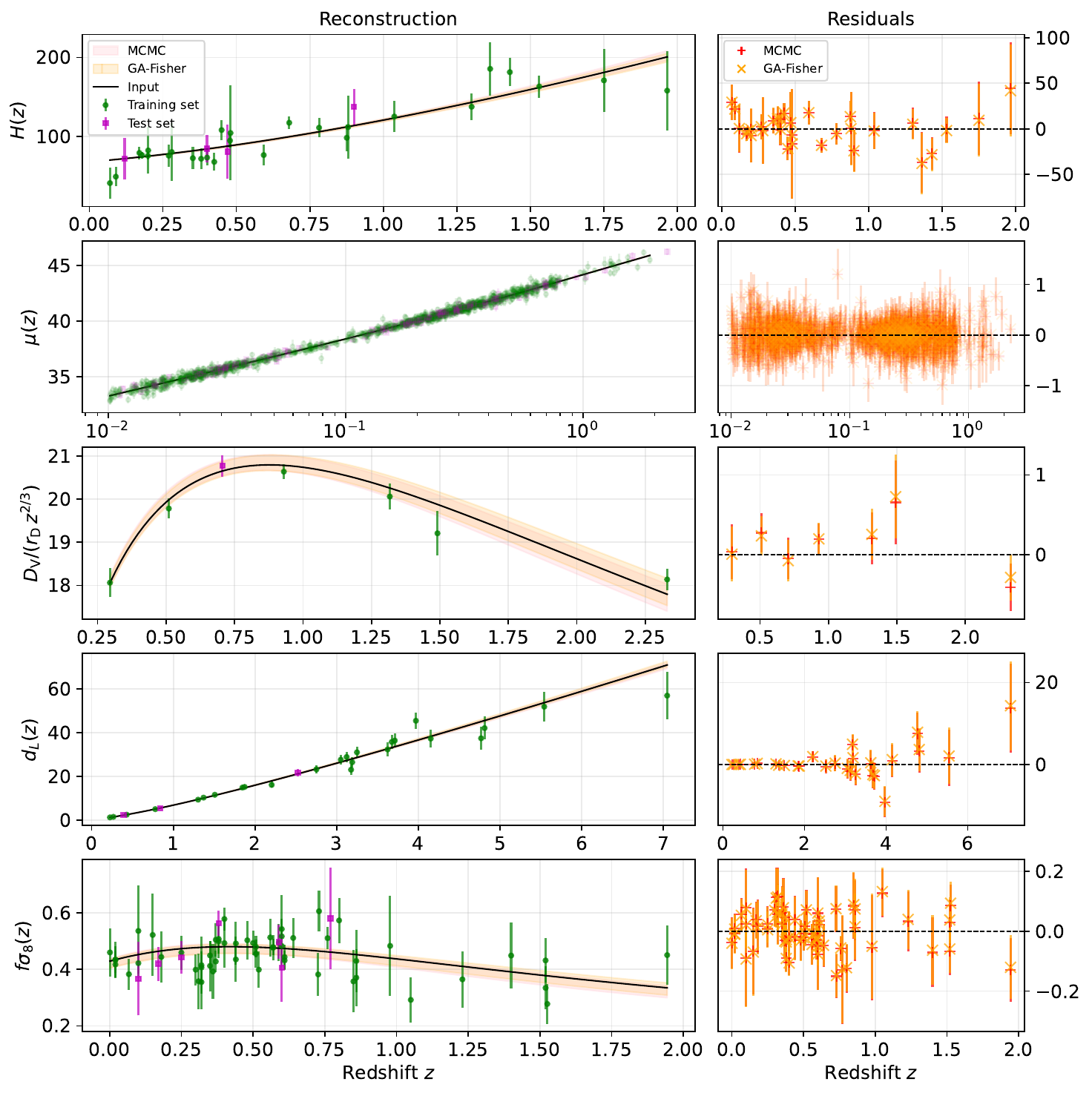}
    \caption{{[Left]} MCMC (red) and GA-Fisher (yellow) best fit curves in $w$CDM in the reference realization (seed/universe 14000605). Black curves show the true cosmology. $2\sigma$ error bands are presented. {[Right] Residuals reconstruction. Residual error bars $\sigma_{\rm res}$ combine the data uncertainty and the reconstruction covariances; e.g., for diagonal covariances, $\sigma_{\rm res}^2 = \sigma_{\rm data}^2 + \sigma_{\rm rec}^2$ where $\sigma_{\rm data}^2$ are data variances and $\sigma_{\rm rec}$ are the diagonals of the reconstruction covariance.}}
    \label{fig:bestfit_ga_fisher}
\end{figure}

This results in MCMC samples stored in \texttt{my\_cl.mcmc\_samples} that can be plotted using standard packages such as \texttt{corner} or \texttt{getdist}. \texttt{cosmo\_learn} has a built-in method \texttt{show\_param\_posterior} that uses the former. In the reference realization (seed 14000605) the MCMC samples are shown in Figure \ref{fig:corner_ga_fisher} together with the corresponding results of GA-Fisher (Section \ref{subsec:geneticalgorithm}).

In addition to corner plots, best fits with error bands can be easily visualized using a method \texttt{show\_bestfit\_curve}, as shown in Figure \ref{fig:bestfit_ga_fisher}, for MCMC and GA-Fisher inferred parameters. These are constructed by resampling the posterior as a Gaussian distribution with a mean and covariance provided by the sampled parameters, and for each sampled point drawing the corresponding observable, making a distribution of functions. Note that this method and the default corner plotter can be used for both MCMC and GA-Fisher, which directly uses the samples stored in the \texttt{cosmo\_learn} instance.

It is also worth noting that the likelihood initialization comes with a default set of priors for the parameters, that users can freely play with by modifying the corresponding python dictionaries, \texttt{prior\_dict} and \texttt{rd\_fid}, as follows:
{\fontsize{8}{10}\selectfont \begin{verbatim}
# priors (flat for H0, Om0, w0, s8, Gaussian for rd_fid)
prior_dict={`H0_min': 0, `H0_max': 100, `Om0_min': 0, `Om0_max': 1, \
            `w0_min': -10, `w0_max': 10, `s8_min': 0.2, `s8_max': 1.5}
rd_fid_prior={`mu': 147.46, `sigma': 0.28}
llprob=lambda x: my_cl.llprob_wcdm(x, prior_dict=prior_dict, rd_fid_prior=rd_fid_prior)
\end{verbatim}}

The likelihood can be extended beyond $w$CDM, should users wish to. A tailored likelihood, e.g., for CPL or axion-like early dark energy, can be setup and given as input \texttt{llprob} to the method \texttt{get\_mcmc\_samples}.

\subsection{GA-Fisher: Genetic Algorithm + Fisher Forecast}
\label{subsec:geneticalgorithm}

Genetic algorithm (GA) is a nature-inspired, population-based optimization strategy that emulates the biological principles of natural evolution \cite{Medel-Esquivel:2023nov, Bernardo:2025flj, Bernardo:2025zbv, DiValentino:2025sru}. It is classified as a metaheuristic method, meaning it does not rely on gradients or derivatives to search for optima. Instead, it heuristically explores the solution space, with the goal of identifying the fittest solution within a population of candidate solutions. GAs are especially powerful in high-dimensional or multimodal problems, where local optima may obscure the global minimum. Under suitable conditions, GAs are known to converge to the global optimum \cite{rudolph1994convergence}.

GA has been applied in various areas of science, including high-energy physics \cite{Akrami:2009hp}, gravitational wave astronomy \cite{Crowder:2006wh}, and increasingly in cosmology. They were introduced to cosmological analysis in \cite{Bogdanos:2009ib} as a model-independent reconstruction tool to alleviate the arbitrariness of dark energy parametrizations. Follow-up studies such as \cite{Nesseris:2010ep, Nesseris:2012tt} established GA as a flexible tool for cosmological parameter estimation and uncertainty quantification. A comprehensive and modern overview of GA for cosmological inference is provided in \cite{Medel-Esquivel:2023nov}. We refer interested readers to \cite{Bernardo:2025zbv, Bernardo:2025flj} for a pedagogical introduction to GA and nature-inspired methods for cosmological parameter estimation.

In the context of cosmology, an individual in the GA population represents a realization of cosmological parameters, e.g., $\left(H_0, \Omega_{m0}, w_0, w_a\right)$. Each individual is encoded as a chromosome, consisting of a set of genes, each corresponding to a parameter (see Figure 2 of \cite{Bernardo:2025zbv}). The GA evolves this population over generations via selection, crossover, and mutation, with the aim of improving the fitness; typically defined via a cost or likelihood function related to observational data.

\begin{figure}[h!]
\centering
\resizebox{0.5\textwidth}{!}{ 
\begin{tikzpicture}[
  node distance = 0.5cm and 1cm,
  every node/.style = {rectangle, draw=black, thick, align=center, minimum width=3.2cm, minimum height=0.9cm, rounded corners=2pt},
  ->, >=latex, thick
]

\node (init) {Initialization};
\node (select) [below=of init] {Selection};
\node (crossover) [below=of select] {Crossover};
\node (elite) [left=of crossover] {Elitism};
\node (mutation) [below=of crossover] {Mutation};
\node (newgen) [below=of mutation] {New Generation};

\draw (init) -- (select);
\draw (select) -- (crossover);
\draw (crossover) -- (mutation);
\draw (mutation) -- (newgen);
\draw (newgen.east) to[out=0,in=270] ++(2,1.5) -- ++(0,3.0) to[out=90,in=0] (init.east);
\draw (select.west) to[out=180,in=90] (elite.north);
\draw (elite.south) to[out=270,in=180] (newgen.west);

\end{tikzpicture}
}
\caption{Flowchart of a GA \cite{CosmoVerse:2025txj, Bernardo:2025zbv}.}
\label{fig:GA_flowchart}
\end{figure}
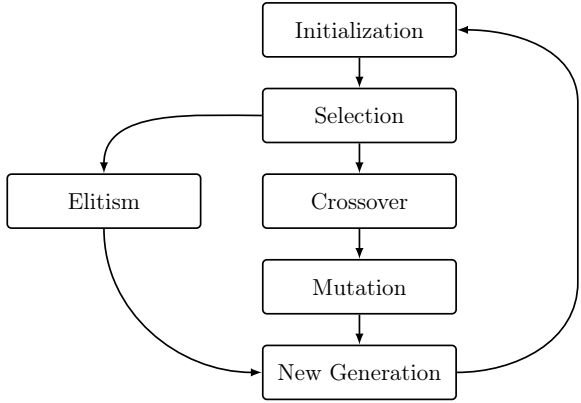

The key ingredients and steps of GA are shown in Table~\ref{tab:GA_ingredients} and Figure~\ref{fig:GA_flowchart}.

\begin{table}[h!]
\centering
\caption{Key components of a GA.}
\label{tab:GA_ingredients}
\renewcommand{\arraystretch}{1.3}
\begin{tabularx}{0.95\textwidth}{>{\bfseries}l X}
Fitness Function & Quantifies how well an individual performs relative to the target objective. \\
Selection & Determines which individuals are chosen to reproduce. Common techniques include roulette-wheel or tournament selection, which favor individuals with higher fitness. \\
Crossover & Combines parent chromosomes to produce offspring, enabling exploration of new regions in the parameter space by mixing genetic material. \\
Mutation & Introduces random genetic changes, to avoid local optima and maintain genetic diversity. \\
Elitism & Preserves top-performing individuals across generations. \\
\end{tabularx}
\end{table}

At its core, GA ranks individuals based on fitness, promotes the fittest to reproduce through crossover and mutation, and ensures survival of the best candidates via elitism. The iterative cycle continues until convergence is achieved or a maximum number of generations is reached.

In cosmological applications, each gene typically represents a parameter such as $H_0$, $\Omega_{m0}$, $w_0$, or $w_a$, while the chromosome is a vector of these parameters. For example, in Chevallier–Polarski–Linder (CPL) dark energy parametrization, the chromosome could be $(H_0, \Omega_{m0}, w_0, w_a)$ \cite{Medel-Esquivel:2023nov}.

Beyond traditional parameter estimation, GA has been deployed in more sophisticated frameworks. In the pioneering works \cite{Bogdanos:2009ib, Nesseris:2010ep, Nesseris:2012tt}, grammatical evolution (GE)—an extension of GA—was used to reconstruct cosmological functions without a fixed parametric form. GA has also been used to eliminate the arbitrariness in kernel choice for Gaussian Process regression \cite{Bernardo:2021mfs}, optimize neural network architectures \cite{Gomez-Vargas:2022bsm}, and improve spectroscopic modeling and model selection \cite{Bainbridge2017a, Bainbridge2017b, Lee2020AI-VPFIT, Webb2021}.

In \texttt{cosmo\_learn}, GA can be performed using a method \texttt{get\_gaFisher\_samples}, e.g., in $w$CDM:
\begin{verbatim}
llprob=lambda x: my_cl.llprob_wcdm(x) # log likelihood function
fitness_func=lambda x: -2*llprob(x) # fitness function
prior_ga=[[0, 100], [0, 1], [-10, 10], [0, 1.5], [145, 149]]
my_cl.get_gaFisher_samples(fitness_func, prior_ga, llprob)
\end{verbatim}
This scans the same parameter space as the built-in MCMC, $(H_0, \Omega_{m0}, w, S_8, r_{\rm D})$, and so the priors [min, max] for each parameter must be given in the same order. Note that our hybrid method applies a Fisher matrix uncertainty estimate on top of optimization, thus requiring an input likelihood probability. Mainly, GA is used to search the global optimum of the likelihood in parameter space, then the likelihood is used to forecast the uncertainty given the Fisher matrix approach. The resulting method is referred to as GA-Fisher.

Figures~\ref{fig:corner_ga_fisher}-\ref{fig:bestfit_ga_fisher} illustrate a representative result of cosmological parameter estimation using the GA-Fisher method in the reference realization 14000605. In this case, GA correctly identifies the best fit parameters corresponding to the maximum likelihood point in the parameter space which agree with the MCMC-derived best fit within $2\sigma$. This agreement extends to the confidence regions: the contours produced by GA-Fisher closely match those from MCMC, both in shape and size. This similarity arises because GA-Fisher uses the likelihood function to estimate uncertainties via the Fisher matrix. Such consistency between the two approaches is expected primarily when the data set is sufficiently constraining, i.e., when multiple cosmological probes such as CC, SNe, RSD, BAO, and GWs are combined. However, it is important to emphasize that the level of agreement between GA-Fisher and MCMC is realization-dependent, driven by the underlying noise in the mock data. While realizations such as the reference case yield good agreement, others may exhibit more significant discrepancies.

The package comes with the following default settings for GA: population size 100, number of generations 1000, parent selection rate 0.3, mutation probability 0.3, uniform crossover rate 0.5. These can be changed to user-preferred values as follows,
\begin{verbatim}
my_cl.ga_params[`max_num_iteration']=100
my_cl.ga_params[`population_size']=50
my_cl.ga_params[`mutation_probability']=0.3
my_cl.ga_params[`elit_ratio']=0.01
my_cl.ga_params[`crossover_probability']=0.5
my_cl.ga_params[`parents_portion'=0.3
my_cl.ga_params[`crossover_type']='uniform'
\end{verbatim}
Lastly the convergence curve (fitness function as a function of generations) can be shown by adding the option \texttt{convergence\_curve=False} when calling the method \texttt{get\_gaFisher\_samples}. 

\texttt{cosmo\_learn} uses the \href{https://github.com/rmsolgi/geneticalgorithm}{\texttt{geneticalgorithm}} \texttt{python} library to perform GA.

\subsection{Gaussian Processes}
\label{subsec:gaussianprocesses}

A Gaussian process (GP) is a generalization of a single Gaussian distribution to a series of Gaussian distributions \cite{10.5555/971143, 10.5555/1162254}. This is normally based on a set of Gaussian points in a training data set. Here, some reconstructed function $H \equiv H(z)$ is defined as a Gaussian process with mean $\mu(z)$ and variance $\sigma^2 (z)$ for any point $z$ within the function domain. The reconstructed profile is formed by considering the relationship that neighbouring points $\tilde{z}$ have on the functional value at $z$. This is quantified through the covariance function ${\rm cov}\left(H(z),H(\tilde{z})\right) = k\left(z, \tilde{z}\right)$, which expresses the strength and separation magnitude that neighbouring points will have on each other. Then, the function can be written as a Gaussian process
\begin{equation}
    H(z) \sim \mathcal{GP}\left(\mu(z), k(z, \tilde{z})\right)\,,
\end{equation}
formed with only the relationship between points in the data set and does not include any physical parameters of some underlying model.

The covariance function can be defined through a kernel which depends on the relationship between neighbouring points of the functional $H$. As an example, we consider the Cauchy kernel; but there are many other possible ways to relate neighbouring points along the function. This is defined through
\begin{equation}
    k(z, \tilde{z}) = \sigma_H^2 \left[\frac{l_H}{(z - \tilde{z})^2 + l_H^2}\right]\,,
\end{equation}
where $\sigma_H$ and $l_H$ are constants that quantify the amplitude and distance at which pairs of points are correlated with each other. For the Cauchy kernel, the point square distance in the denominator means that the further points are from each other the weaker their covariance will be. This decay is polynomial which can be more realistic than other options such as exponential scenarios. However, in all cases, it is the non-parametric constants that characterize the resulting functional behavior for $H(z)$. It is important to emphasize that while the $\sigma_H$ and $l_H$ parameters do relate an expected statistical behavior of the points along the $H(z)$, there is no information embodied about any underlying physical model.

Taking some set of points $Z^* = \{z^*\}$, one can define a mean $\mu^*$ and kernel $\left[K(Z,Z)\right]_{ij} = k(z_i^*, z_j^*)$ from which the function values $H^* = H(z_i^*)$ are described by
\begin{equation}
    H^* \sim \mathcal{N}\left(\mu^*, K(Z^*, Z^*)\right)\,,
\end{equation}
where $\mathcal{N}$ refers to the GP as evaluated at the specific points in $Z^*$. While the values of $H^*$ form a normal distribution, its values at specific points $z_i^*$ are determined by the correlations embodied in the kernel function and respective mean $\mu^*$. The function $H^*$ is very arbitrary and is only set by setting values for the mean and the kernel constants, or statistical hyperparameters. Another way to make this determination is to use observational data to set these parameter values and then reconstruct the function at other points where there is no data. 

One can consider observations information $Z = \{z_i\}$ which have known functional values $(z_i, \overline{H}(z_i))$, where $\overline{H}(z_i)$ must have Gaussian uncertainties. These points form part of a normal distribution or GP evaluated at specific points at which data is known
\begin{equation}
    \overline{H}(z_i) \sim \mathcal{N}\left(\mu(z_i), k(Z,Z) + C\right)\,,
\end{equation}
where $C$ is the covariance matrix which accounts for correlations between the data. For scenarios where there are no correlations, the covariance matrix takes on the form $C = {\rm Diag} \left( \sigma_i^2 \right)$. Together with GP, this data can be used to reconstruct information at the unknown points $Z^*$ using observations at $Z$ points. This is achieved by first combining both distributions
\begin{equation}
    \begin{pmatrix}
        \overline{H} \\
        H^*
    \end{pmatrix}
    \sim \mathcal{N} \left(
    \begin{pmatrix}
        \mu \\
        \mu^*
    \end{pmatrix}\,,
    \begin{pmatrix}
        k(Z,Z) + C & k(Z,Z^*) \\
        k(Z^*, Z) & k(Z^*, Z^*)
    \end{pmatrix}
    \right)\,.
\end{equation}
It is through this combination of observational data and target reconstruction points, that the reconstructed function's profile can be obtained. This is achieved through a supervised training process in which the marginal likelihood
\begin{equation}
    p\left(\overline{H}\vert Z,\,\sigma_H, l_H\right) = \int p\left(\overline{H}\vert H, Z\right) p\left(H\vert Z, \sigma_H, l_H\right)\, dH\,,
\end{equation}
is maximized, where the $H$ represents GP function values, and $Z$ are points where observational points exist. Notice that $Z^*$ does not feature in this integral, meaning that the points at which the reconstruction takes place are not part of the marginalization process. Once the GP hyperparameters are constrained, the locations at which reconstruction is to take place can be performed.

Given a Gaussian prior $H\vert Z, \sigma_H, l_H \sim \mathcal{N}\left( \mu, k(Z,Z) \right)$ and taking the observational data to be a Gaussian distribution at every point $\overline{H}\vert H \sim \mathcal{N}\left( H, C\right)$, results in the log marginal likelihood
\begin{align}
    \ln \mathcal{L} &= \ln p\left(\overline{H}\vert Z,\,\sigma_H, l_H\right)\nonumber\\
    &= -\frac{1}{2} \left(\overline{H} - \mu\right)^T \left[k(Z,Z) + C\right]^{-1} \left(\overline{H} - \mu\right) - \frac{1}{2} \ln|k(Z,Z) + C| - \frac{n}{2} \ln 2\pi\,,
\end{align}
which accounts for covariances between the observational data set, and provides a likelihood through which to optimize the nonphysical hyperparameters. For most scenarios, the mean can be set to zero without loss of generality provided this is done before the hyperparameters are fit. The GP hyperparameters will be valued accordingly to account for all information obtained from the training data \cite{Seikel2012, Shafieloo:2012ht, Seikel:2013fda}. In this way, cosmological data can be used to realize reconstructed diagrams for various cosmological parameters, including the Hubble diagram \cite{Shafieloo:2012ht, Colgain:2021ngq,Yennapureddy:2017vvb,Seikel2012,Seikel:2013fda, Benisty:2020kdt, Belgacem:2019zzu,Moore:2015sza,Canas-Herrera:2021qxs, Briffa:2020qli,Cai:2019bdh, LeviSaid:2021yat, Cai:2015zoa, Reyes:2021owe, Wang:2017jdm, Gomez-Valent:2018hwc, Zhang:2018gjb, Mukherjee:2020vkx, Aljaf:2020eqh, Li:2019nux, Liao:2019qoc, Busti:2014aoa, Cai:2015pia, Renzi:2020fnx, Bernardo:2021cxi}.

\begin{figure}[h!]
    \centering
    \includegraphics[width=0.9\textwidth]{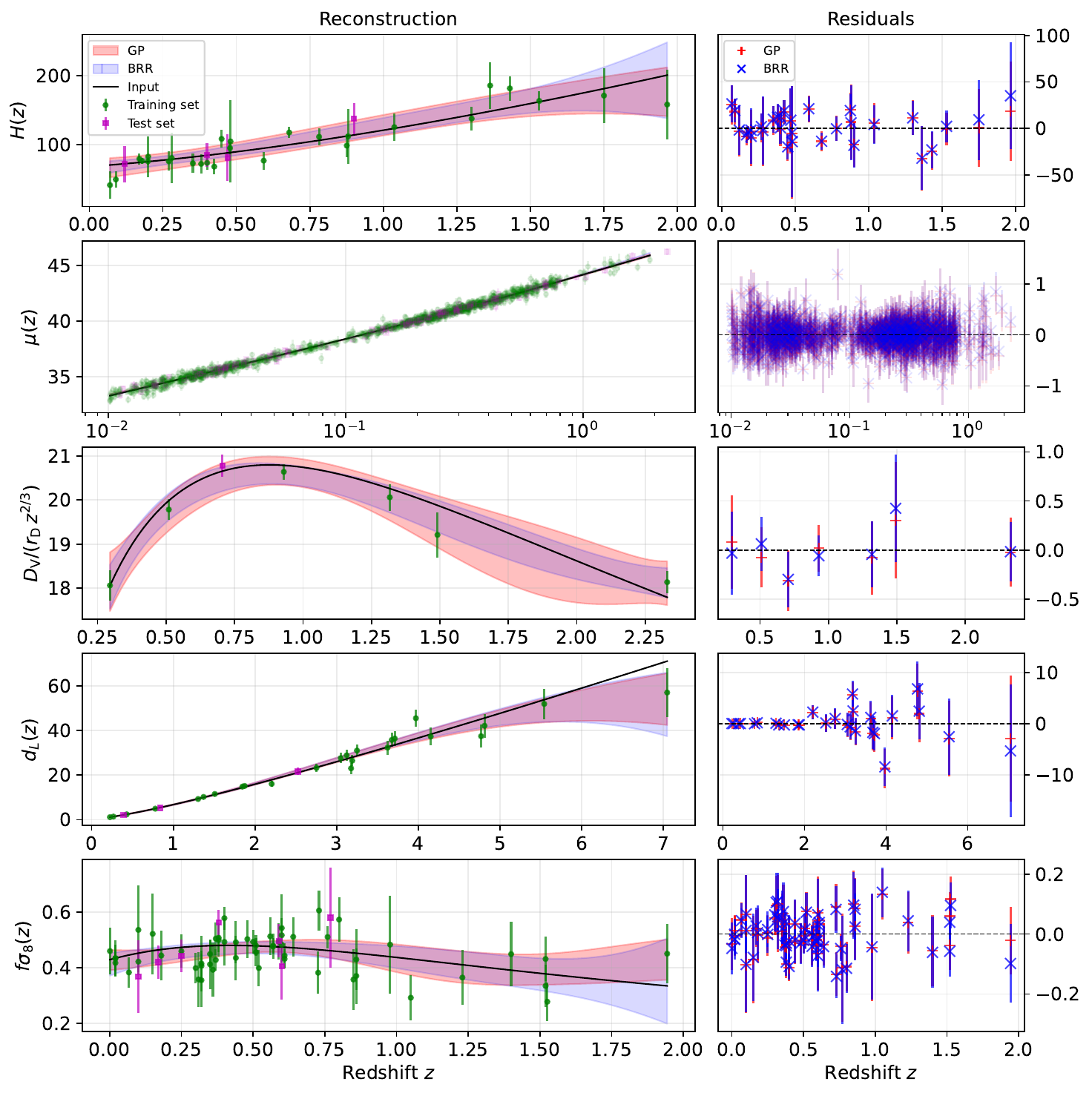}
    \caption{{[Left]} GP (red) and 
    BRR (blue) reconstructions of the cosmological functions in mock data 14000605. Black curves show the true cosmology. $2\sigma$ error bands are presented. {[Right] Residuals reconstruction. Residual error bars $\sigma_{\rm res}$ combine the data uncertainty and the reconstruction covariances; e.g., for diagonal covariances, $\sigma_{\rm res}^2 = \sigma_{\rm data}^2 + \sigma_{\rm rec}^2$ where $\sigma_{\rm data}^2$ are data variances and $\sigma_{\rm rec}$ are the diagonals of the reconstruction covariance.}}
    \label{fig:gp_brr}
\end{figure}

In \texttt{cosmo\_learn}, a GP can be trained with a single command following the initialization of a mock data set:
\begin{verbatim}
    my_cl.train_gp()
\end{verbatim}
This function fits a separate GP model for each cosmological probe present in \texttt{mock\_data}. The training uses a default of \texttt{n\_restarts\_optimizer = 10} to help avoid local minima in the hyperparameter space.

By default, the package employs a radial basis function (RBF or Gaussian) kernel, but alternative kernels can be specified via the \texttt{kernel\_key} option in \texttt{train\_gp}. Available choices include \texttt{Matern}, \texttt{RationalQuadratic} (which generalizes the Cauchy kernel), \texttt{ExpSineSquared}, and \texttt{DotProduct}. In all cases, the hyperparameter initial values and bounds follow \texttt{scikit-learn}'s defaults.

GP regression is non-parametric and highly flexible, allowing for smooth interpolation of the data with uncertainty estimates. For CC, BAO, RSD, and GW data sets, the model is trained in linear redshift space, while for SNe data, it operates in $\log_{10}(z)$ to better accommodate the steep slope at low redshift. A comparison of default GP and BRR reconstructions using the mock reference cosmology is shown in Figure~\ref{fig:gp_brr}.

\texttt{cosmo\_learn} performs GP regression using \texttt{scikit-learn}’s \texttt{GaussianProcessRegressor} module \cite{scikit-learn}.

\subsection{Bayesian Ridge Regression}
\label{subsec:bayerianridge}

Bayesian regression techniques offer a probabilistic approach to linear modeling, where regularization parameters are inferred from the data rather than fixed a priori. This allows the model to adapt its complexity to the data at hand. Unlike classical ridge regression, where the regularization strength $\alpha$ is manually set, Bayesian regression treats it as a hyperparameter drawn from a prior distribution \cite{bishop2006pattern, neal2012bayesian}.

Ridge regression, also known as L2 regularization, is a linear regression technique used to mitigate multicollinearity in multiple regression data. Multicollinearity occurs when two or more independent variables are highly correlated, leading to unstable coefficient estimates. In ridge regression, a regularization term $R$ is added to the ordinary least squares (OLS) cost function:
\begin{equation}
R = \alpha \, ||w||^2
\end{equation}
where $\alpha$ is the regularization hyperparameter and $w$ is the coefficient vector. The regularization term penalizes large coefficients, helping to reduce model variance and prevent overfitting.

Bayesian Ridge Regression builds on this by placing a Gaussian prior over the coefficients $w$, with zero mean and precision $\lambda$:
\begin{equation}
    p(w \,|\, \lambda) = \mathcal{N}(0, \lambda^{-1} I) \,.
\end{equation}
The likelihood is modeled as a Gaussian distribution of the targets $y$ around the linear model $Xw$ with noise precision $\alpha$:
\begin{equation}
    p(y \,|\, X, w, \alpha) = \mathcal{N}(Xw, \alpha^{-1} I) \,.
\end{equation}
To make the model fully Bayesian, Gamma priors are placed over both $\alpha$ and $\lambda$, the conjugate priors for Gaussian precision. The hyperparameters are estimated by maximizing the log marginal likelihood (also known as evidence approximation). This leads to an automatic trade-off between model complexity and data fidelity.

As with GP regression, Bayesian Ridge Regression (BRR) training in \texttt{cosmo\_learn} can be performed with a single command:
\begin{verbatim}
    my_cl.train_brr()
\end{verbatim}
This routine trains an independent BRR model for each cosmological probe available in \texttt{mock\_data}, using polynomial basis functions of degree \texttt{n\_order}, which defaults to 3. The polynomial order determines the model’s capacity to capture non-linear trends in the data: higher orders yield more flexible fits but may risk overfitting.

BRR extends linear regression by introducing priors over the model coefficients, allowing for automatic regularization. The model internally applies inverse-variance weighting based on the provided uncertainties and initializes the hyperparameters for the coefficient and noise precisions with \texttt{alpha\_init = 1.0} and \texttt{lambda\_init = $10^{-3}$}, respectively. These priors are iteratively updated during training to maximize the marginal likelihood.

For the SNe data set, training is carried out in $\log_{10}(z)$ instead of $z$ to better resolve the structure of the Hubble diagram at low redshift. Reconstructions using the default BRR settings are shown alongside Gaussian Process regression in Figure~\ref{fig:gp_brr} for direct comparison.

\texttt{cosmo\_learn} uses \texttt{scikit-learn}’s \texttt{BayesianRidge} estimator \cite{scikit-learn} to perform BRR, following the formulation of Refs.~\cite{mackay1992bayesian, tipping2001sparse}.

\subsection{Neural Nets}
\label{subsec:neuralnets}

Artificial neural networks (ANN) can be used as another method by which to reconstruct cosmological evolution parameters in an accurate and fast manner \cite{2015arXiv151107289C}. ANN structures are composed of input and output layers through which interface with the input data and output predictions can be respectively made. A series of parameters connecting hidden layers can then be optimized to mimic real training data relationships. Each layer is composed of a number of neurons that connect to neurons in the layers preceding or proceeding that layer.

\begin{figure}[h]
	\begin{center}
		
		\begin{tikzpicture}[scale=0.8]
		
		\foreach \i in {1,...,\inputnum}
		{
			\node[circle, 
			minimum size = 7mm,
			fill=green!50] (Input-\i) at (0,-\i) {};
		}
		
		\foreach \i in {1,2,3}
		{
			\node[circle, 
			minimum size = 7mm,
			fill=blue!40,
			yshift=(\hiddennum-\inputnum)*5 mm
			] (Hidden1-\i) at (2.5,-\i) {};
		}
		
		\node(dots) at (2.5,-2){\vdots};
		
		\node[circle, 
		minimum size = 7mm,
		fill=blue!40,
		yshift=(4-\inputnum)*5 mm
		] (Hidden1-5) at (2.5,-4) {};

		\foreach \i in {1,...,\hiddennum}
		{
			\node[circle, 
			minimum size = 7mm,
			fill=blue!40,
			yshift=(\hiddennum-\inputnum)*5 mm
			] (Hidden1-\i) at (2.5,-\i) {};
		}

		\foreach \i in {1,2,3}
		{
			\node[circle, 
			minimum size = 7mm,
			fill=blue!40,
			yshift=(\hiddennum-\inputnum)*5 mm
			] (Hidden2-\i) at (4.5,-\i) {};
		}
		
		\node(dots) at (4.5,-2){\vdots};
		
		\node[circle, 
		minimum size = 7mm,
		fill=blue!40,
		yshift=(4-\inputnum)*5 mm
		] (Hidden2-5) at (4.5,-4) {};

		\foreach \i in {1,...,\hiddennum}
		{
			\node[circle, 
			minimum size = 7mm,
			fill=blue!40,
			yshift=(\hiddennum-\inputnum)*5 mm
			] (Hidden2-\i) at (4.5,-\i) {};
		}

		\foreach \i in {1,2,3}
		{
			\node[circle, 
			minimum size = 6.5mm,
			fill=blue!40,
			yshift=(\hiddennum-\inputnum)*5 mm
			] (Hidden3-\i) at (6.5,-\i) {};
		}
		
		\node(dots) at (6.5,-2){\vdots};
		
		\node[circle, 
		minimum size = 7mm,
		fill=blue!40,
		yshift=(4-\inputnum)*5 mm
		] (Hidden3-5) at (6.5,-4) {};

		\foreach \i in {1,...,\hiddennum}
		{
			\node[circle, 
			minimum size = 7mm,
			fill=blue!40,
			yshift=(\hiddennum-\inputnum)*5 mm
			] (Hidden3-\i) at (6.5,-\i) {};
		}
		
		\foreach \i in {1,...,\outputnum}
		{
			\node[circle, 
			minimum size = 7mm,
			fill=red!50,
			yshift=(\outputnum-\inputnum)*5 mm
			] (Output-\i) at (9.5,-\i) {};
		}
		
		\foreach \i in {1,...,\inputnum}
		{
			\foreach \j in {1,...,\hiddennum}
			{
				\draw[->, shorten >=1pt, color=black!50] (Input-\i) -- (Hidden1-\j);	
			}
		}

		\foreach \i in {1,...,\hiddennum}
		{
			\foreach \j in {1,...,\hiddennum}
			{
				\draw[-, shorten >=1pt, color=black!50] (Hidden1-\i) -- (Hidden2-\j);	
			}
		}

		\foreach \i in {1,...,\hiddennum}
		{
			\foreach \j in {1,...,\hiddennum}
			{
				\draw[-, shorten >=1pt, color=black!50] (Hidden2-\i) -- (Hidden3-\j);	
			}
		}
		
		\foreach \i in {1,...,\hiddennum}
		{
			\foreach \j in {1,...,\outputnum}
			{
				\draw[->, shorten >=1pt,  color=black!50] (Hidden3-\i) -- (Output-\j);
			}
		}
		
			\draw[<-, shorten >=1pt] (Input-1) -- ++(-1,0)
			node[left]{Input 1};
            \draw[<-, shorten >=1pt] (Input-2) -- ++(-1,0)
			node[left]{Input 2};
            \draw[<-, shorten >=1pt] (Input-3) -- ++(-1,0)
			node[left]{Input 3};
   
		\draw[->, shorten >=1pt] (Output-1) -- ++(1,0)
		node[right]{Output 1};
        \draw[->, shorten >=1pt] (Output-2) -- ++(1,0)
		node[right]{Output 2};

        \draw [dashed] (1.5,0) -- (1.5,-4.2);
        \draw [dashed] (7.5,0) -- (7.5,-4.2);
		
		\end{tikzpicture}
		\vskip -0.3cm
		
  Input Layer~~~~~~~~~~~~~~~~~~~~~~~~~~~~~~~~~~Hidden Layers~~~~~~~~~~~~~~~~~~~~~~~~~~~~~~~~~~~~~Output Layer 
		
	\end{center}
    \caption{An illustrative example of a general NN structure is shown where three inputs are translated to two outputs through a series of three hidden layers.} 
\label{fig:ANN_structure}
\end{figure}
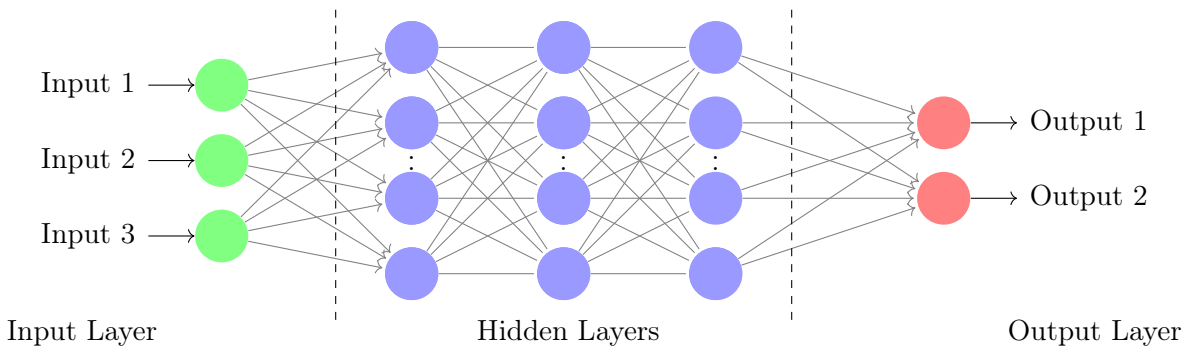

As an illustrative example, we show Figure~\ref{fig:ANN_structure} where a series of three hidden layers structure the NN neurons and which has three inputs together with two outputs. In this way, an input signal would traverse the entirety of the network architecture in order to produce an output. The NN is constructed such that a linear transformation (composed of linear weights and biases) is employed for each of the different layers.

The neurons in the NN are constructed out of an activation function which can be used to model complex relationships in the data when there are many neurons. In this work, and in many works, the Exponential Linear Unit (ELU) \cite{2015arXiv151107289C} is used which is specified by
\begin{equation}
    f(x) = 
  \begin{cases} 
   {x} & \text{if } x>0 \\
   {\alpha(e^x-1)} & \text{if } x \leq 0
  \end{cases}\,,
\end{equation}
where $\alpha$ is a strictly positive hyperparameter that moderates the value at which the ELUs saturate for negative inputs, which are then set to unity. The action of this function is to shift negative inputs closer to unity while leaving positive ones unchanged. Through the entirety of the NN, the combinations of the linear transformations and activation functions produce a vast array of hyperparameters that specify the NN responses. These can be optimized through training when their values are optimized for a specific set of training data. For each iteration during the process of training, the difference between the NN predictions and the ground truth in the training data is characterized by a so-called loss function which should be minimized. In the NN implementations that follow, we adopt Adam's algorithm \cite{2014arXiv1412.6980K} which employs an improvement of the traditional gradient descent method which has been modified to accelerate convergence.

The simplest of these loss functions is the absolute difference between the training and predicted outputs which is then summed for each redshift. This is called the L1 loss function, and is defined as
\begin{equation}
    {\rm L1} = \sum_i |\Upsilon_{\rm train}(z_i) - \Upsilon_{\rm pred}(z_i)|\,,
\end{equation}
where $\Upsilon_{\rm train}(z)$ and $\Upsilon_{\rm pred}(z)$ are observed and NN predicted values of cosmological parameter $\Upsilon$ at observation redshifts $z$. This is the most similar to the log-likelihood function for uncorrelated data for MCMC methods. Other popular choices of loss functions include the mean squared error (MSE) loss function, which minimizes the squared differences between the training and predicted outputs, as well as the smoothed L1 (SL1) loss function, which softens the L1 absolute difference for similar values of outputs. It is at the level of the loss function that the complexities inherent in the training data are adopted by the NN. In our work, we adopt an {L1} loss function since it is the most versatile and has shown itself to be the most efficient in training NN architectures in the cosmological regime. 

\begin{figure}[h!]
    \centering
    \includegraphics[width=0.9\textwidth]{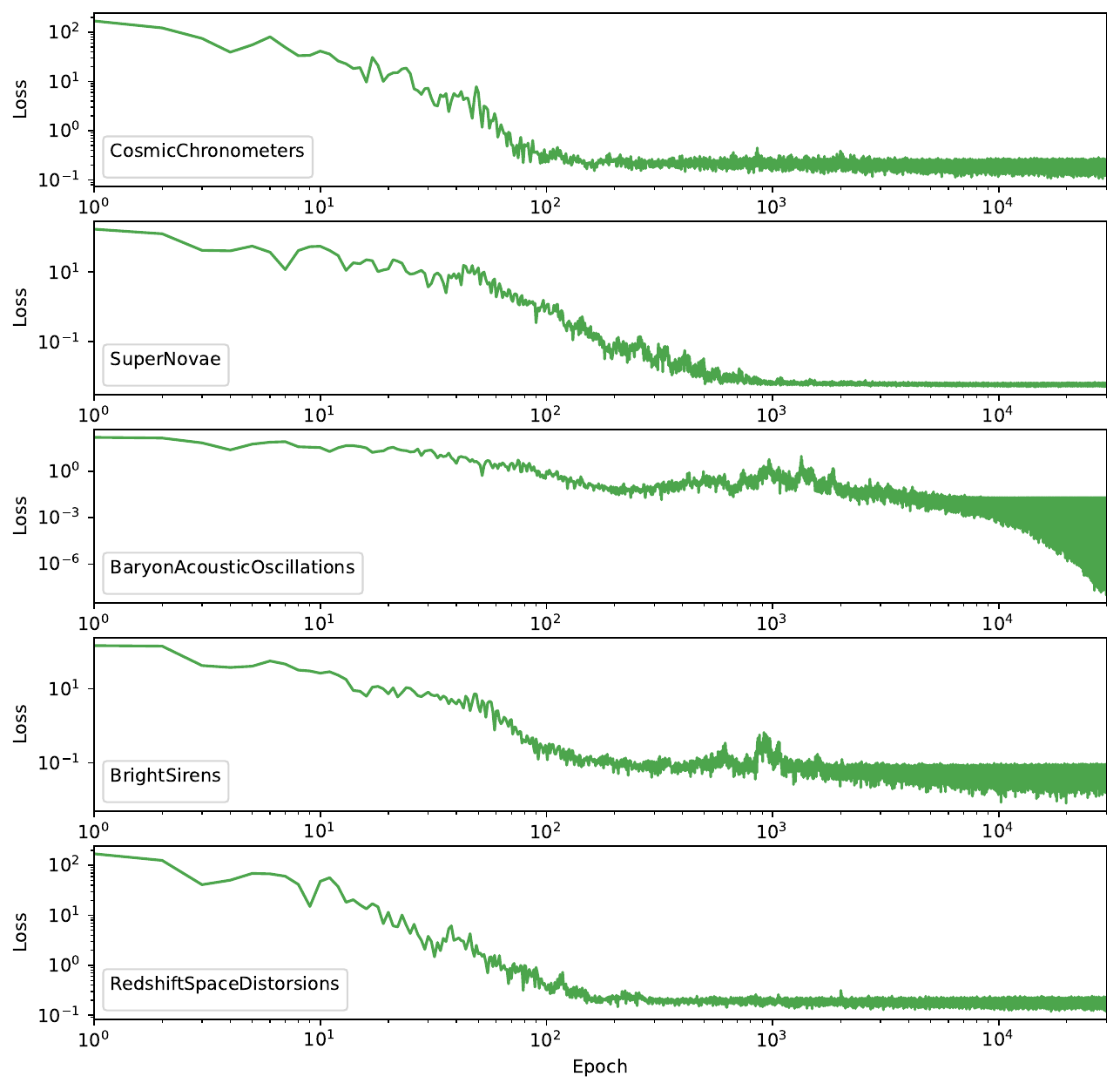}
    \caption{Loss function in the ANN reconstruction of the cosmological functions in reference universe 14000605.}
    \label{fig:ann_loss}
\end{figure}

\begin{figure}[h!]
    \centering
    \includegraphics[width=0.9\textwidth]{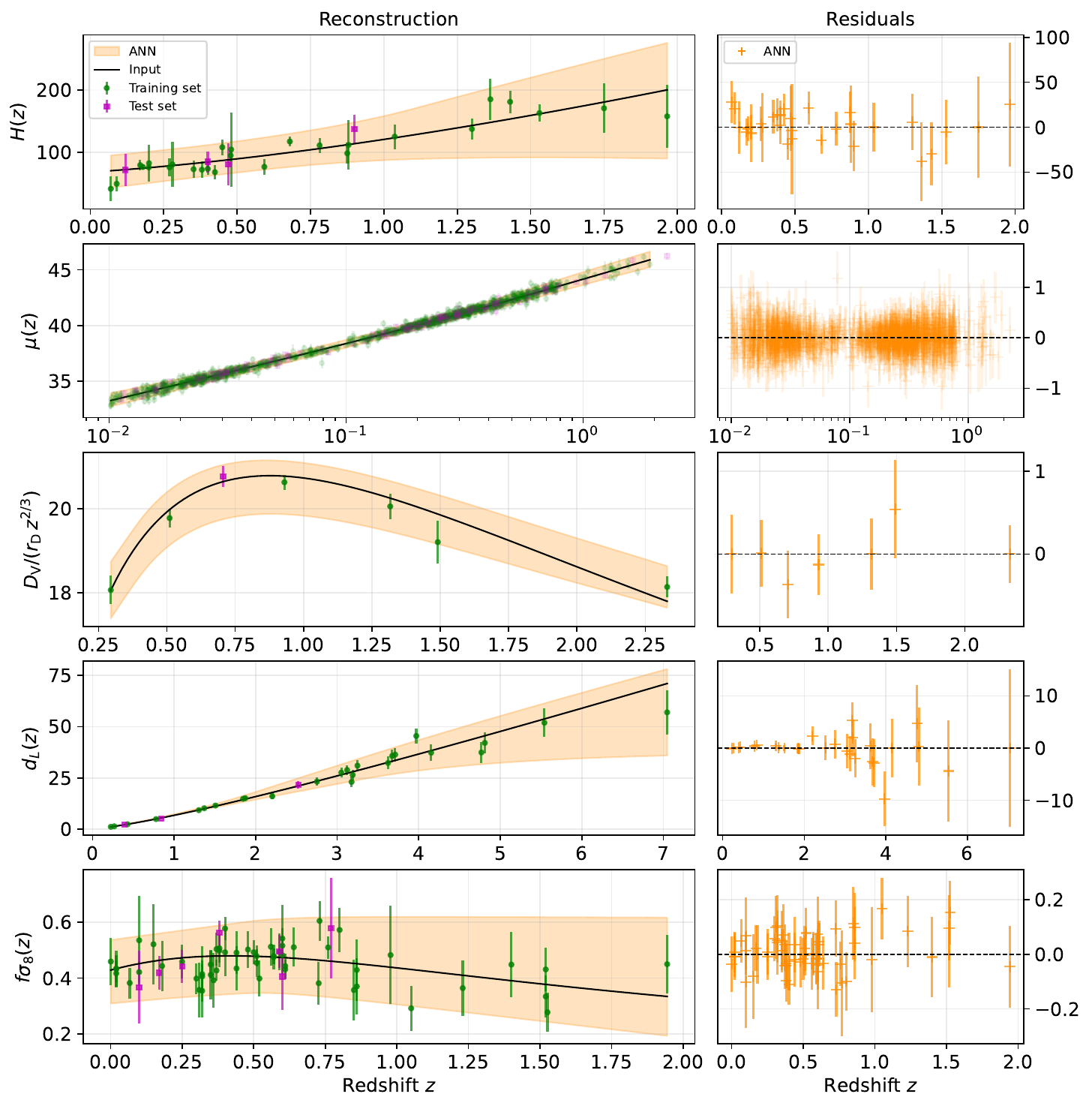}
    \caption{{[Left]} ANN reconstruction of the cosmological functions in reference universe 14000605. Black curves show the true cosmology. $2\sigma$ error bands are presented. {[Right] Residuals reconstruction. Residual error bars $\sigma_{\rm res}$ combine the data uncertainty and the reconstruction covariances; e.g., for diagonal covariances, $\sigma_{\rm res}^2 = \sigma_{\rm data}^2 + \sigma_{\rm rec}^2$ where $\sigma_{\rm data}^2$ are data variances and $\sigma_{\rm rec}$ are the diagonals of the reconstruction covariance.}}
    \label{fig:ann}
\end{figure}

Trained NN can be useful in reconstructing the evolution of various cosmological parameters such as reconstructing the expansion velocity and growth of large scale structures as in Refs.~\cite{Wang:2020sxl,Gomez-Vargas:2021zyl,Dialektopoulos:2021wde}. This removes the reliance on a fiducial model to extract points that do not coincide with observational redshifts. In Ref.~\cite{Benisty:2022psx}, NN methods were used to test whether the absolute magnitude of type 1a supernovae had a significant aspect of evolution. NN implementations were also used to reconstruct specific classes of cosmological models such as in Refs.~\cite{Dialektopoulos:2023jam,Mukherjee:2022yyq} in which a Monte Carlo approach was taken for the derivatives of the reconstructed parameters. Recently, there has also been work on addressing the known problem of incorporating covariance matrices into the training process of NN implementations as in Ref.~\cite{Dialektopoulos:2023dhb}.

ANN training can be performed with \texttt{cosmo\_learn} in two lines, initializing the architecture for each data set and training using a default architecture:
\begin{verbatim} 
    my_cl.init_ann()
    my_cl.train_ann()
\end{verbatim}
The results are shown in Figures \ref{fig:ann_loss} and \ref{fig:ann}. {The convergence of the Loss function does appear to be slow for the individual data sets, which may be caused by several factors including the nuances of the underlying data, the complexity of the calculated cosmological parameters, as well as the configuration of the Universe in question. On the other hand, the convergence rate is not far beyond other comparable universe settings.}

In our implementation, we employ the neural network package \texttt{ReFANN} \cite{Wang:2019vxv} for function reconstruction. As with GP and BRR approaches, each data set is treated independently. For all reconstructions, we adopt the default \texttt{ReFANN} architecture, which consists of a single hidden layer with 4096 neurons and uses the hyperparameter configuration \texttt{`rec\_1'}, characterized by the absence of batch normalization and dropout; use the option \texttt{hp\_model=`rec\_2'} for batch normalization. The number of epochs (default is 30000) used for training can be changed using the \texttt{iteration} option.

All models are compiled using the Adam optimizer~\cite{2014arXiv1412.6980K} and trained with the L1 loss function. {The regularization terms of \texttt{ReFANN} were specifically optimized for cosmological function reconstruction \cite{Wang:2019vxv}, and we found that no further tuning was required to reach our illustrative purposes. However, the effect of regularization terms on reconstruction is non-trivial and may be data- and probe-dependent. This is an avenue for data-driven exploration through \texttt{cosmo\_learn}, with results potentially deserving of a dedicated, systematic treatment, which we reserve for a separate work.}

For the SNe data set, as in the GP and BRR methods, redshifts are logarithmically rescaled before training the ANN to better capture features across a broad dynamic range. The full architecture used for each reconstruction can be printed upon initialization by setting the option \texttt{show\_summary=`True'}. To display training outputs, the option \texttt{print\_info=`True'} can be used during initialization.

After training, the final architectures along with the evolution of training losses are stored in a dictionary accessible via \texttt{my\_cl.ANN\_dict}. For instance, the method \texttt{show\_ann\_loss} (which produced Figure \ref{fig:ann_loss}) retrieves the training history from this dictionary for each data set. This modular structure enables flexible integration of heterogeneous cosmological data sets within a unified ANN-based reconstruction framework.

The results of ANN reconstruction on the reference mock data is shown in Figures \ref{fig:ann_loss}-\ref{fig:ann}. The ANN architecture can be customized for specific purposes by modifying the dictionary \texttt{ANN\_dict} after initialization via \texttt{init\_ANN} and before training via \texttt{train\_ANN}. 

The package \texttt{cosmo\_learn} internally leverages \texttt{reFANN}~\cite{Wang:2019vxv}, which is built on \texttt{pytorch}~\cite{Paszke:2019xhz}, to carry out ANN training and cosmological reconstructions.

\subsection{Benchmarks}
\label{subsec:benchmarks}

{A timely effort afforded by a singular unifying framework for cosmological simulation, inference and reconstruction, such as \texttt{cosmo-learn}, is comparing the performance of old and each new statistical framework that is introduced into cosmological analysis. This is work in progress. Nonetheless, we can assess the performance of the different methods in \texttt{cosmo-learn} with simplified metrics, based on the computation time and the residuals obtained after subtracting the best fit or mean reconstruction to the data points. For this we rely to the SNe Ia mock data set as the densest of our data sets and the reference universe 14000605; this avoids having to complicate this initial benchmark with dealing with the residuals carrying different units for different observables. The result is presented in Table \ref{tab:benchmark}.}

\begin{table}[h!]
\centering
\caption{{Comparison on SNe Ia mock data (reference universe 14000605; processor: Apple M4); $t$ is training time in seconds, Mean, RMS are the residuals mean and root-mean-square values over the training and the test sets, $\chi^2_\mathrm{red}$ is the reduced chi-squared, $\chi^2_{\rm red} = \sum \left( ( y_{\rm rec} - y_{\rm data} ) / \sigma_{\rm res} \right)^2 / N$, where $y_{\rm rec}$, $y_{\rm data}$ are the means of the reconstruction and the data at each redshift, and $\sigma_{\rm res} ^2$ is the combined variance of the reconstruction and the data.}}
\label{tab:benchmark}
\renewcommand{\arraystretch}{1.3}
\begin{tabular}{l r rrr rrr}
\hline\hline
 & & \multicolumn{3}{c}{Training set} & \multicolumn{3}{c}{Test set} \\
\cmidrule(lr){3-5}\cmidrule(lr){6-8}
Method & $t$ (s)
       & Mean & RMS & $\chi^2_\mathrm{red}$
       & Mean & RMS & $\chi^2_\mathrm{red}$ \\
\hline
GP                &  30.593 & $-$0.0016 & 0.2387 & 0.9614 & $-$0.0185 & 0.2223 & 1.0030 \\
BRR      &   0.001 & $-$0.0017 & 0.2382 & 0.9608 & $-$0.0171 & 0.2212 & 0.9978 \\
ANN      &  67.362 & $-$0.0018 & 0.2384 & 0.4939 & $-$0.0179 & 0.2216 & 0.4841 \\
MCMC      &  30.605 & $-$0.0046 & 0.2381 & 0.9600 & $-$0.0189 & 0.2200 & 0.9876 \\
GA-Fisher &   9.836 & $-$0.0076 & 0.2383 & 0.9619 & $-$0.0226 & 0.2210 & 1.0037 \\
\hline
\end{tabular}
\end{table}

{This shows that the effectiveness of each method in minimizing the residuals during training, but also when predicting cosmological realizations in the test set. Firstly, the training times of the different methods vary significantly; in order of increasing computation speed, ANN takes the most time, GP and MCMC take up roughly the same time, GA-Fisher, and BRR. We emphasize that these were obtained with the default configurations of each method in \texttt{cosmo-learn} and that the performance of each can be improved with further tuning hyperparameters. In MCMC this depends mostly on the number of dimensions of the parameter space, in GA-Fisher the mutation rate and number of function evaluations, in GP mostly on the number of data points since matrix inversion is an ${\cal O}(N^3)$ procedure. On the other hand, ANN's training efficiency depends on its architecture (number of layers, neurons, loss function) and BRR on the degree and type of basis functions used to construct the design matrix. BRRs speed advantage is derived through analytic tractability: hyperparameters are iteratively estimated with expectation-maximization and the computational cost is ${\cal O}(N p^2 + p^3)$ for $p$ basis functions (with $p=4$ utilized as the default in \texttt{cosmo-learn}'s BRR module). Clearly, scalability is an advantage for analyzing the rapidly increasing space of cosmological data.}

{However, scalability is only next in priority to the actual performance in terms of reconstruction of a statistical method. The mean, RMS and reduced chi-squared values in Table \ref{tab:benchmark} show us a glimpse of the reconstruction performances reflected in residuals with respect to the SNe Ia mock data in the reference universe. Overall, all methods achieve comparatively small residuals mean and RMS, and $\chi^2_{\rm red}$ of order unity on both the training and test sets, indicating statistically consistent reconstructions. The performance of ANN in terms of the reduced chi-squared metric is noteworthy. It is worth noticing that the generally negative mean obtained across methods is a feature of the underlying realization (in the reference seed 14000605); in any realization, it is expected that the data may sit a hair above or below the fiducial curves. A rigorous comparison of these methods should be carried out across multiple mock realizations and observational probes, using a set of unbiased metrics; this is left for future work.}

\section{Conclusions}
\label{sec:conclusions}

In this paper, we have introduced the \texttt{python}-based package \texttt{cosmo\_learn}, a flexible and user-friendly toolkit for simulating cosmological data and performing data-driven inference. The software supports realistic noise modeling for various cosmological probes, including cosmic chronometers, supernovae, baryon acoustic oscillations, redshift space distortions, and gravitational waves. We have verified the consistency of the simulated data with the input cosmological model through residuals and parameter inferences. Additionally, we have demonstrated the core functionalities of several built-in learning and inference tools, including MCMC, genetic algorithm, Gaussian processes, Bayesian ridge regression, and artificial neural networks. Emphasis was placed on maintaining simplicity and adaptability throughout the development process to offer users an intuitive and extensible platform for exploring cosmology through simulation and machine learning.

Future developments of the package aim to incorporate additional cosmological data sets, such as cosmic microwave background measurements and big bang nucleosynthesis priors, to allow for more consistent and comprehensive modeling of initial conditions. Existing modules will be updated as improved data or models become available. While the current implementation uses a constant dark energy equation of state for which both background evolution and linear perturbations have analytic solutions, the framework can be readily extended to support dynamical dark energy models \cite{Wang:2001ht, Cardenas:2014jya, Pan:2019gop, Grandon:2021nls}. The current learning framework is also designed to be extensible, making it straightforward to implement methods such as Approximate Bayesian Computation \cite{Bernardo:2021mfs, Bernardo:2022ggl, Bernardo:2022pyz} and other emerging techniques that offer complementary statistical perspectives on cosmological tensions. We encourage contributions from the community and welcome modifications, extensions, and forks of our open-source codebase, which will remain publicly available and free to use.

With \texttt{cosmo\_learn}, we hope to support both teaching and research efforts at the interface of cosmology, simulation, and machine learning.

\bmhead{Acknowledgements}

RCB was supported by an appointment to the JRG Program at the APCTP through the Science and Technology Promotion Fund and Lottery Fund of the Korean Government, and was also supported by the Korean Local Governments in Gyeongsangbuk-do Province and Pohang City. RCB thanks the Institute of Physics, Academia Sinica, Taiwan for the hospitality that enabled the completion of this work. This article is also based upon work from COST Action CA21136 Addressing observational tensions in cosmology with systematics and fundamental physics (CosmoVerse) supported by COST (European Cooperation in Science and Technology). DG and VHC acknowledge partial support from Centro de Física Teórica de Valparaíso CEFITeV.  JLS would also like to acknowledge funding from ``Xjenza Malta'' as part of the ``Technology Development Programme'' DTP-2024-014 (Cosmic Learning) Project. RCB and GCB thank participants of the external program APCTP-GW2025 [or APCTP2025-E05] held at Academia Sinica, Taipei, Taiwan for fruitful discussions.

\bmhead{Data availability}
No new data were generated in this work.

\bmhead{Code availability}
The python code \texttt{cosmo-learn} is publicly available in GitHub: \href{https://github.com/reggiebernardo/cosmo_learn}{https://github.com/reggiebernardo/cosmo\_learn}.










\begin{appendices}






\end{appendices}




\end{document}